# A Statistical Review of Light Curves and the Prevalence of Contact Binaries in the Kuiper Belt

Mark R. Showalter[a*], Susan D. Benecchi[b], Marc W. Buie[c], William M. Grundy[d], James T. Keane[e], Carey M. Lisse[f], Cathy B. Olkin[c], Simon B. Porter[c], Stuart J. Robbins[c], Kelsi N. Singer[c], Anne J. Verbiscer[g], Harold A. Weaver[f], Amanda M. Zangari[c], Douglas P. Hamilton[h], David E. Kaufmann[c], Tod R. Lauer[i], D. S. Mehoke[f], T. S. Mehoke[f], J. R. Spencer[c], H. B. Throop[b], J. W. Parker[c], S. Alan Stern[c], and the New Horizons Geology, Geophysics, and Imaging Team.

[a] SETI Institute, Mountain View, 189 Bernardo Ave., CA 94043, USA.
[b] Planetary Science Institute, 1700 East Fort Lowell, Suite 106, Tucson, AZ 85719, USA.
[c] Southwest Research Institute, 1050 Walnut St., Suite 300, Boulder, CO 80302, USA.
[d] Lowell Observatory, 1400 W. Mars Hill Rd., Flagstaff, AZ 86001, USA.
[e] California Institute of Technology, 1200 E. California Blvd, Pasadena, CA 91125, USA.
[f] Johns Hopkins University Applied Physics Laboratory, 11100 Johns Hopkins Rd., Laurel, MD 20723, USA.
[g] Department of Astronomy, University of Virginia, 530 McCormick Rd., Charlottesville, VA 22904, USA.
[h] Department of Astronomy, University of Maryland, College Park, MD 20742, USA.
[i] National Optical Astronomy Observatory, 950 N. Cherry Ave., Tucson, AZ 85719, USA.
*Correspondence: mshowalter@seti.org

## Abstract

We investigate what can be learned about a population of distant Kuiper Belt Objects (KBOs) by studying the statistical properties of their light curves. Whereas others have successfully inferred the properties of individual, highly variable KBOs, we show that the fraction of KBOs with low amplitudes also provides fundamental information about a population. Each light curve is primarily the result of two factors: shape and orientation. We consider contact binaries and ellipsoidal shapes, with and without flattening. After developing the mathematical framework, we apply it to the existing body of KBO light curve data. Principal conclusions are as follows.(1) When using absolute magnitude *H* as a proxy for the sizes of KBOs, it is more accurate to use the maximum of the light curve (minimum *H*) rather than the mean. (2) Previous investigators have noted that smaller KBOs tend to have higher-amplitude light curves, and have interpreted this as evidence that they are systematically more irregular in shape than larger KBOs; we show that a population of flattened bodies with uniform proportions, independent of size, could also explain this result. (3) Our method of analysis indicates that prior assessments of the fraction of contact binaries in the Kuiper Belt may be artificially low. (4) The pole orientations of some KBOs can be inferred from observed changes in their light curves over time scales of decades; however, we show that these KBOs constitute a biased sample, whose pole orientations are not representative of the population overall. (5) Although surface topography, albedo patterns, limb darkening, and other surface properties can affect individual light curves, they do not have a strong influence on the statistics overall. (6) Photometry from the Outer Solar System Origins Survey (OSSOS) survey is incompatible with previous results and its statistical properties defy easy interpretation. We also discuss the promise of this approach for the analysis of future, much larger data sets such as the one anticipated from the upcoming Vera C. Rubin Observatory.

## 1. Introduction

The New Horizons spacecraft completed the first close reconnaissance of a small Kuiper Belt object (KBO) in January 2019. The target body, (486958) Arrokoth (also identified as 2014 $MU_{69}$) was revealed to be a contact binary, which resulted from the merger of two flattened, spheroidal components (McKinnon et al., 2020; Spencer et al., 2020; Stern et al., 2019; see Fig. 1). Arrokoth has a near-circular, near-ecliptic orbit 44 AU from the Sun, placing it at the edge of the cold classical Kuiper Belt (Petit et al. 2011); this suggests that it probably formed in situ very early in the history of the Solar System, and has been largely unmodified since.

Many KBO formation models predict that binaries, and often contact binaries, are a likely outcome (Fraser et al., 2017; Nesvorný and Vokrouhlický, 2019; Nesvorný et al., 2010, 2018). The detailed images of Arrokoth provide dramatic support for this prediction, and further demonstrate the importance of understanding the full distribution of body shapes within the Kuiper Belt. The most widely available method for constraining the shapes of KBOs is through rotational light curve observations (e.g., Lacerda and Luu, 2003). Opportunities for more direct determinations of the shape, via a close spacecraft flyby (Stern et al., 2019) or via multi-chord stellar occultation observations (Buie et al., 2020; Ortiz et al., 2017), require substantially greater resources and are therefore much more limited.

The light curves of many main belt asteroids have been successfully “inverted” to determine their shapes in considerable detail (Ďurech et al., 2010; Ostro and Connelly, 1984; Ostro et al., 1988). However, these techniques are not currently applicable to KBOs for three key reasons. First, the inversion technique requires the viewing of a body from multiple aspect angles. Because KBOs are distant and move slowly, so far only a few objects, (139775) 2001 $QG_{298}$ (Lacerda, 2011) and (20000) Varuna (Fernández-Valenzuela et al., 2019), have long enough measurement baselines to begin this sort of analysis. Second, due to their small size and great distance, few KBOs have been sampled with sufficient regularity and signal-to-noise (SNR) for inversion techniques to be applicable. Third, all inversion techniques assume that the body is convex; the technique will therefore always lead to unreliable conclusions if applied to a body with major concavities such as, for example, Arrokoth (Harris and Warner, 2020).

The alternative technique is forward-modeling: to define a family of shapes and determine which of these provide good fits to the available photometric data. Such shape determinations are never unique, but they can be very informative, particularly when used as tests of theoretical expectations. For example, “dog bone”-shaped figures have been proposed as a way to understand the light curves of some Solar System bodies (Descamps 2015, 2016). The simplest plausible shape for any small body is a triaxial ellipsoid, which is expected to produce a roughly sinusoidal light curve as it rotates. To date, however, numerous KBOs have revealed distinctly non-sinusoidal light curves, which are inconsistent with ellipsoids but which are, in fact, consistent with models for contact binaries (Lacerda, 2011; Sheppard and Jewitt, 2002; Thirouin and Sheppard, 2017, 2018, 2019; Thirouin et al., 2017).

The interpretation of a light curve requires another important consideration: what is the orientation of the rotation pole? Here, again, Arrokoth provides a useful illustration. Although it is highly elongated, it showed no detectable light curve variations from Earth orbit (Benecchi et al., 2019; Stern et al., 2019) or during New Horizons' approach in December 2018 (Zangari et al., 2019). Part of the explanation is that the viewpoints of Earth and the New Horizons spacecraft were both close to Arrokoth's rotation anti-pole. At low phase angles, a view from exactly along the pole would result in a flat light curve, independent of the shape. As a corollary, it is incorrect to assume that a body is "round" if all we know about it is that its light curve is flat.

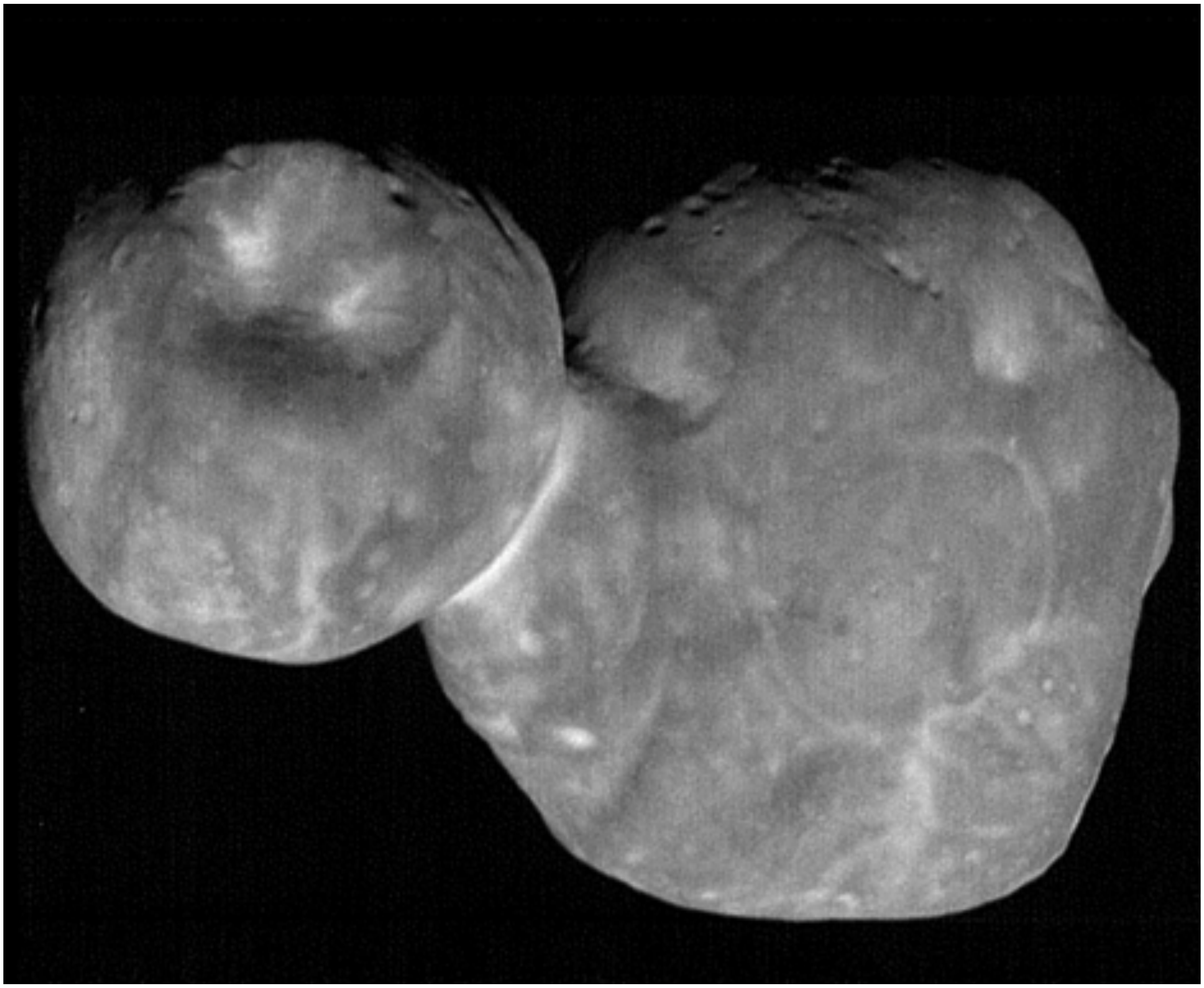

**Figure 1**. Arrokoth, a cold classical contact binary, as imaged by the New Horizons spacecraft. The long axis is ~ 36 km (Spencer et al., 2020; Stern et al., 2019).

Although a single light curve has no unique interpretation, it is possible to constrain the shapes and orientations of an ensemble of KBOs from the statistical properties of their light curves. Here we present a simple thought experiment. Suppose we were to observe that the light curves of a particular family of KBOs are all flat. Any one body might have its pole pointed toward or away from Earth, but it is unlikely that they all do. Therefore, it would be reasonable to conclude that these bodies are predominantly spherical or oblate—shapes that produce flat light curves regardless of their aspect angle.

In this paper, we explore the inferences that one can potentially make about an ensemble of KBOs given the statistical properties of its light curves. We focus, in particular, on the key differences between two simple shape models: triaxial ellipsoids and contact binaries. In Section 2, we model individual hypothetical light curves based on shape, orientation, and surface properties. In Section 3, we relate the statistical properties of the light curves to the statistical properties of the bodies. In Section 4, we compare the available data to these models and draw some preliminary conclusions. Finally, in Section 5, we discuss the implications of this work, including its potential application to the much larger collections of KBO photometry that we anticipate in the near future.

## 2. Models for Rotational Light Curves

### *2.1 Idealized Shapes*

Figure 2 shows light curves for two idealized, rotating shapes: a contact binary consisting of two identical, uniform spheres, and a prolate ellipsoid with proportions $a{:}b{:}c$ = 2:1:1, where $a$, $b$, and $c$ are the radii along the three principal axes and $a \geq b \geq c$. These two shapes have the same physical cross sections when viewed along all three axes, but their light curves are very different. The distinctive broadened peak in brightness and the narrow, "V"-shaped trough are the key distinguishing characteristic of the light curves of contact binaries, be they asteroids (Benner et al., 2006; Tedesco, 1979) or KBOs (Lacerda, 2011; Sheppard and Jewitt, 2002; Thirouin and Sheppard, 2017, 2018, 2019; Thirouin et al., 2017).

We have derived these light curves by constructing numerical 3-D models and then illuminating and rotating them. We define the orientation of each shape model by two angles: $\theta$ is the rotation angle of the body about its shortest axis, such that $\theta$ = 0 presents the largest cross-section to the observer; $\phi$ is the pole angle or aspect angle, measured from the rotation axis to the direction of the observer.

In our models, self-shadowing is included for surface topography and between the lobes, but multiple scattering is not.[1] We model the dependence of surface reflectivity $R$ on lighting and viewing geometry using a Lommel-Seeliger law (Hapke, 2012; Lumme and Bowell, 1981;

[1] The single-scattering approximation is widely used for numerous reasons. Multiple scattering only becomes important for bodies that have (1) high albedos, (2) markedly concave shapes, and (3) are observed at high phase angles. We explain: (1) Low albedo bodies do not reflect much sunlight at all, and therefore one surface region cannot reflect much sunlight onto another. (2) Obviously, convex bodies cannot reflect sunlight onto themselves. (3) The regions of a body's surface that might be self-illuminated are those facing toward other, sunlit surfaces; these same regions are generally pointed away from the Sun, and so are not visible to the observer when the phase angle is low. Accounting for multiple scattering can also substantially increase computational complexity and time, necessitating the tracing of individual light rays as they bounce off two or more surfaces. Although Arrokoth is distinctly concave, Spencer et al. (2020) did not attempt to model multiple scattering when calibrating their reflectance map (their Fig. 1b).

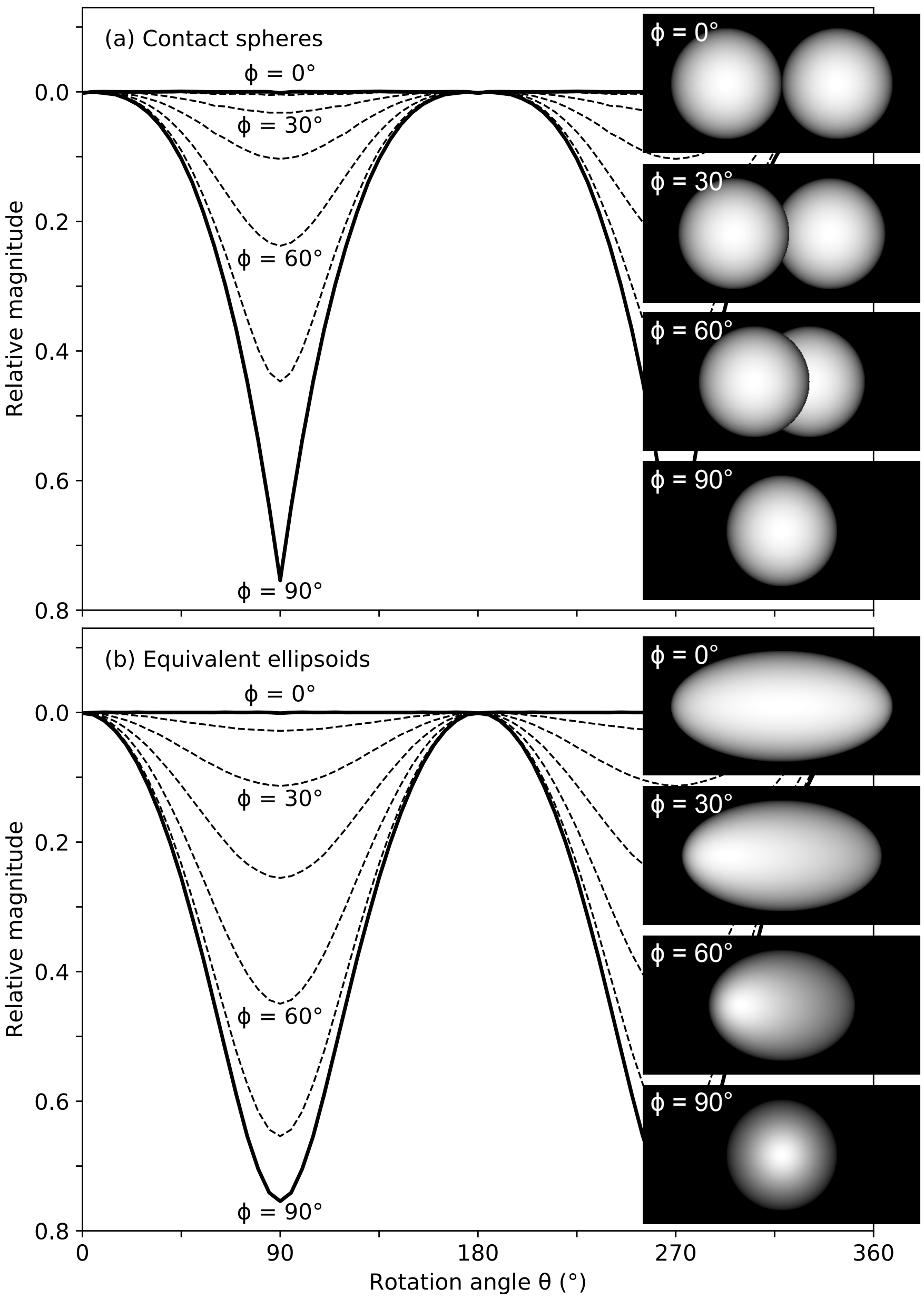

**Figure 2**. Simulated rotational light curves for two idealized shapes, illustrating key differences between contact binaries and ellipsoids. A body comprising two equal spheres in contact (a) shows deep, narrow brightness minima and broader maxima. A 2:1:1 ellipsoid (b) shows light curves that are more sinusoidal in nature. Each curve encompasses 360° of rotation for a different orientation of the pole: $\phi = 0°$ corresponds to a view along the rotation pole, whereas $\phi = 90°$ is for a view perpendicular to the pole. Dotted lines show the intermediate light curves at 15° steps in $\phi$. The inset figures show each shape's appearance on the sky at minimum brightness ($\theta = 90°$), for the specified value of $\phi$ (rotated toward the right). The vertical axis is in units of magnitudes, but values increase downward so that brightness increases upward in the curves.

Seeliger, 1884), where $R \propto \mu_0 / (\mu + \mu_0)$. Here, $\mu_0$ is the cosine of the incidence angle and $\mu$ is the cosine of the emission angle. This law is appropriate for most solid bodies; we explore alternative models below in Section 2.4. However, note that most of the sample images we present in this paper, such as the insets in Fig. 2, use a Lambert law, $R' \propto \mu_0$; this was chosen because we have found that the resulting limb darkening makes these figures easier to interpret. In the limit where phase angle $\alpha = 0$, the angles $\mu$ and $\mu_0$ become equal, so $R$ is constant, meaning that the light curve depends only on the projected cross-section of the body. In this limit, both sets of light curve models can be determined analytically; see the appendix for details.

The inset images in Fig. 2a illustrate a distinctive trait of contact binaries: the lobe in front obscures only a tiny fraction of the lobe behind it unless our point of view is very close to the long axis. This is the reason for the narrow, downward troughs in the model light curves of Fig. 2a. It also illustrates why the depth of the light curve's dip is much smaller for the contact binary (Fig. 2a) than for the ellipsoid (Fig. 2b) at intermediate values of $\phi$. As a specific example, when $\phi = 60°$, the contact binary has a peak-to-peak amplitude $\Delta m$ of 0.25 mag, whereas the ellipsoid has $\Delta m \approx 0.45$ mag. This, as we have since learned, was a secondary contributor to the "missing" light curve of Arrokoth (Benecchi et al., 2019; Stern et al., 2019; Zangari et al., 2019). The angle between the New Horizons approach vector and the pole was ~ 141° (Spencer et al., 2020). From this viewpoint, if Arrokoth had been a ~ 2:1 prolate ellipsoid rather than a contact binary, New Horizons would have observed and easily measured $\Delta m \approx 0.2$ mag.

Figure 3 illustrates this phenomenon more generally. It compares the light curves for contact binaries and equivalent-area ellipsoids using a broader range of body shapes. We have created models of contact binaries comprising two touching spheres in which the radius of the secondary ranges from 40% to 100% that of the primary (cf. Fig. 2). For each ratio, we show the same curves for an "equivalent" prolate ellipsoid, which presents the same cross-sectional area along its three principal axes. Compared to the ellipsoids, contact binaries always have broader peaks and narrower, steeper drops toward their brightness minima. However, as Figs. 2a and 3a reveal, the sharp, "V"-shaped minimum only exists for $\phi$ near 90° and for radius ratios near 100%. For smaller ratios, the minima are truncated when the smaller lobe falls entirely in front of, or behind, the larger lobe; this produces flat brightness minima in the light curves (Fig. 3a) and flat maxima in the dependence of $\Delta m$ on $\phi$ (Fig. 3b). Even for equal-sized lobes, the light curve minima are smooth, not sharp, for $\phi < 90°$ (Fig. 2a). In these cases, it could be more difficult to distinguish the light curves of contact binaries from those of ellipsoids. This is consistent with the fact that, to date, only KBOs with well-sampled light curves and $\Delta m \gtrsim 0.4$ mag have been interpreted as contact binaries (Lacerda, 2011; Sheppard and Jewitt, 2002; Thirouin and Sheppard, 2017, 2018, 2019; Thirouin et al., 2017). Other contact binaries could be escaping our notice simply by having less ideal pole directions or lobes of unequal size.

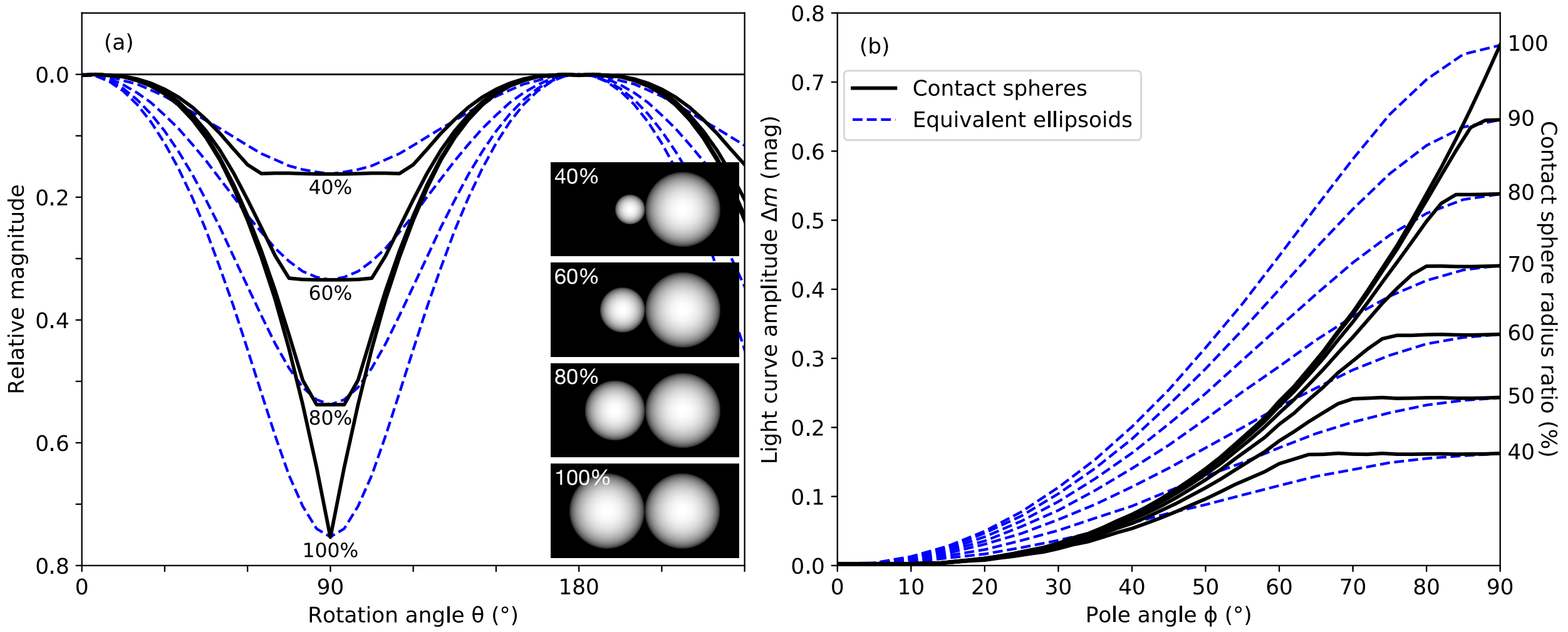

**Figure 3**. Light curve models for contact binaries with varying lobe sizes are compared to those of ellipsoids with equivalent dimensions. Inset images show binaries in which the smaller lobe has a radius between 40% and 100% that of the larger. (a) The rotational light curves for these models are compared to those of equivalent ellipsoids, all assuming $\phi$ = 90°. The curve repeats every 180°, but we show 240° of rotation so that both extrema are easier to see. (b) Dependence of the peak-to-peak light curve amplitude $\Delta m$ on the orientation of the pole. Labels at right indicate the radius of the smaller sphere relative to that of the larger for the contact binary. The flat areas near θ = 90° in panel a and $\phi$ = 90° in panel b arise because the smaller lobe falls entirely in front of or behind the larger lobe.

### *2.2 Effects of Topography*

The shapes discussed above are perfectly idealized. Irregularities, surface topography, and albedo patterns will all modify light curves. To explore these effects, we have performed a variety of alterations to our 3-D models.

To model the role of topography, we performed a tessellation of the sphere into 80 near-equilateral triangles by starting with the vertices of an icosahedron and then subdividing each face into four sub-triangles. For ellipsoidal models, we then scaled the three axes as needed. Finally, we randomized the radius of each vertex by replacing its radius $r_k$ with $r'_k = r_k + q\, r_0\, x_k$, where $x_k$ is a normally-distributed random variable, $r_0$ is the reference radius of the body (the radius for a sphere; the intermediate radius $b$ for an ellipsoid) and $q$ defines the amplitude of the distortion. Inset images in Fig. 4 show examples of these distorted bodies using $q = 0.06$.

We performed ten independent realizations of each random shape. Figure 4a shows the resulting light curves and Figure 4b shows $\Delta m$ vs. pole angle $\phi$, as in Fig. 3. In general, these numerical models closely track the results obtained for the idealized spheres and ellipsoids. The most notable change is a small (< 0.1 mag) upward shift in $\Delta m$ for the two-sphere model below $\phi \approx 45°$; this is not unexpected given how small $\Delta m$ would be otherwise. Overall, these models increase $\Delta m$ by 0.03 ± 0.03 magnitudes.

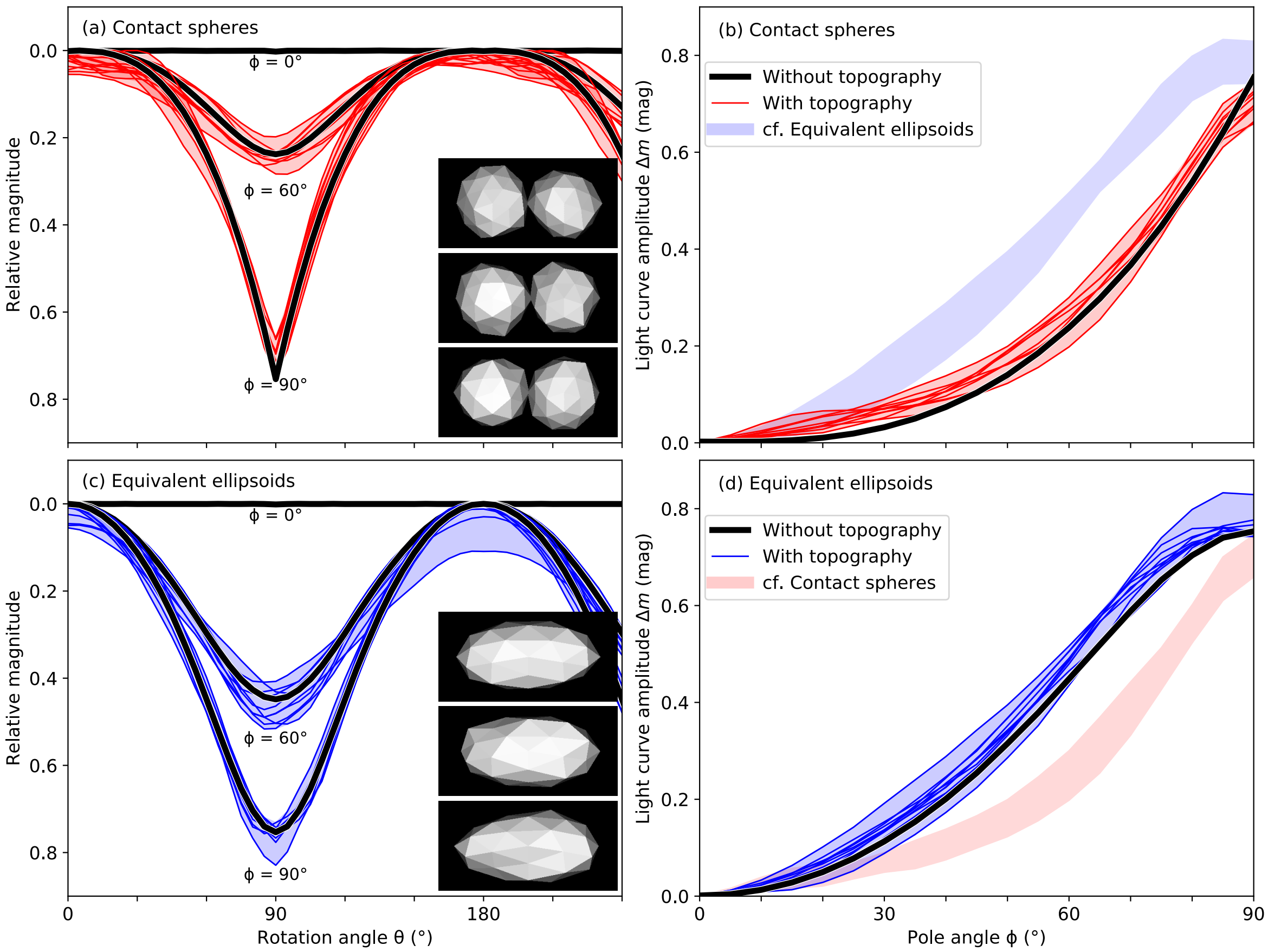

**Figure 4.** Simulations of the effects of irregular shapes. The original body shapes were touching spheres of equal size (panels a,b) and 2:1:1 ellipsoids (c,d). Left panels show rotational light curves for pole angles $\phi$ of 0°, 60°, and 90°; the patterns repeat every 180°, but we show 240° of rotation so that the peaks of the light curves are easier to examine. Right panels show light curve amplitudes $\Delta m$ vs. $\phi$. Inset images show three examples of each shape. In each panel, thin lines show the results for ten random realizations of each shape model, and light shading fills in the zone between the extreme values. Heavy lines show modeling results for the original, undistorted shapes (cf. Fig. 3). The shaded zone of panel d is duplicated in panel b, and vice-versa, to make it easier to compare the two sets of models. In most cases, shape irregularities change light curves by < 0.1 mag. Amplitudes Δm increase under most circumstances, but they can sometimes decrease for certain shapes and orientations.

Spencer et al. (2020) report that the dominant topographic feature on Arrokoth is a 0.51 km depression, probably an impact crater, on the smaller lobe. The depth amounts to ~ 7% of that lobe's 7-km radius. Elsewhere, Arrokoth is smooth and only lightly cratered. For comparison, the 3-D models rendered for Fig. 4 are quite extreme, having ubiquitous peaks and depressions with root-mean-square amplitudes of 6%. If Arrokoth is at all representative of the population overall, then we can infer that localized topography is not likely to be a large contributor to the amplitudes of most KBO light curves; typical contributions will be at the level of a few hundredths of a magnitude or less.

### *2.3 Effects of Albedo Variations*

For exploring the implications of surface albedo variations, we used the same tessellation as described above in Section 2.2. However, instead of changing the radii of the vertices, we assigned a random albedo to the triangular patch associated with each face (Fig. 5, inset images). Albedos were drawn from a uniform distribution between 0.5 and 1. Although these particular values are higher than one would expect for a KBO, the derived light curves are applicable to any body with bright and dark patches that differ in reflectivity by up to a factor of two. Figure 5 shows the modified light curves produced by ten different realizations of each albedo pattern. These albedo variations systematically increase $\Delta m$ by 0.03 ± 0.03 mag, although changes as large as 0.1 mag are sometimes observed.

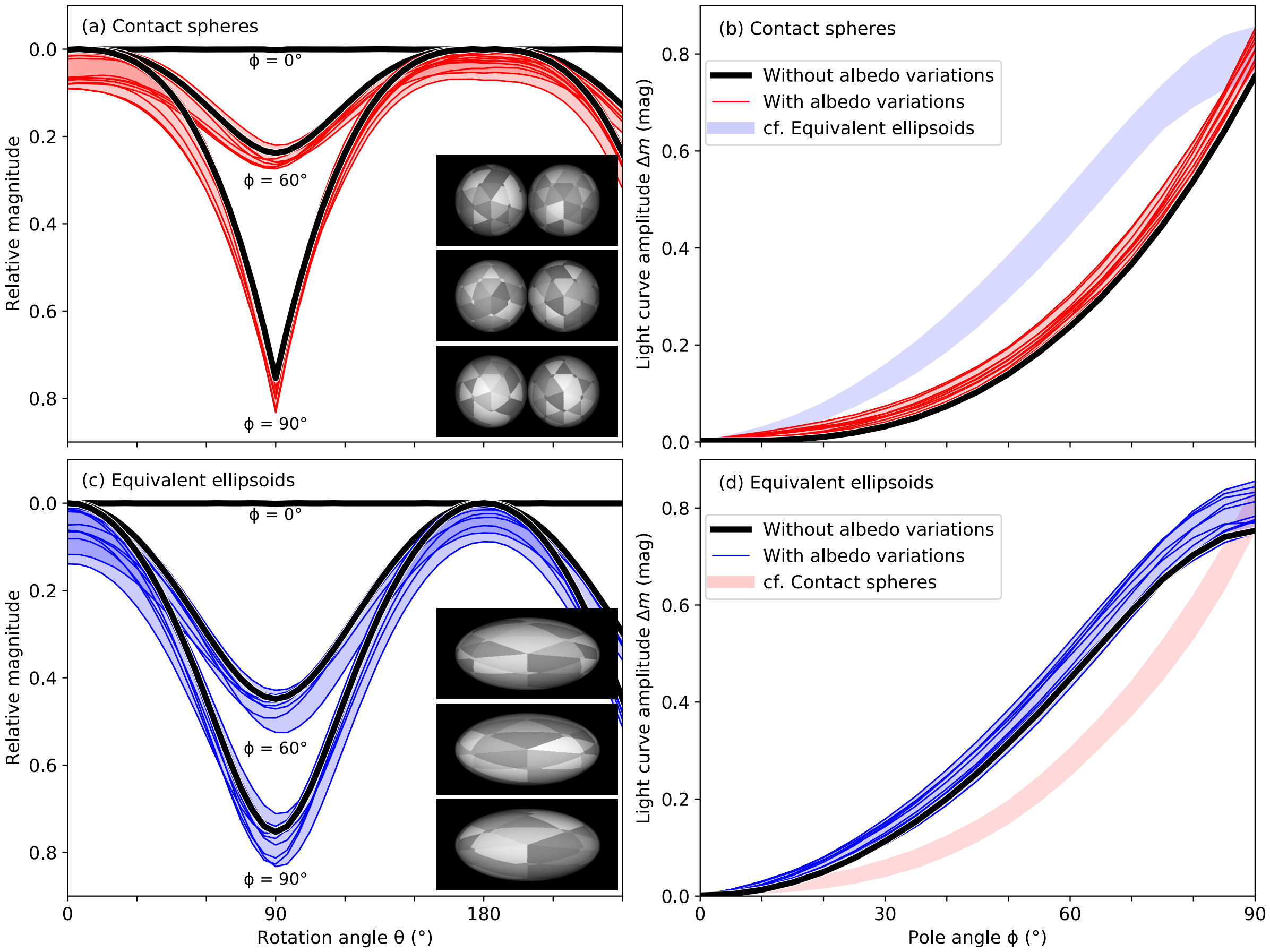

**Figure 5**. Simulations of the effects of albedo patterns for touching spheres (panels a,b) and 2:1:1 ellipsoids (c,d). The surface of each body has been divided into triangles, which have been randomly assigned albedo values between 0.5 and 1. Left panels show rotational light curves for pole angles $\phi$ of 0°, 60°, and 90°; right panels show light curve amplitudes $\Delta m$ vs. $\phi$. Inset images show three examples of each albedo pattern. In each panel, thin lines show the results for ten random realizations of each pattern, and light shading fills in the zone between the extreme values. Heavy lines show modeling results for the original, constant-albedo models (cf. Fig. 3). The shaded zone of panel d is duplicated in panel b, and vice-versa, to make it easier to compare the two sets of models. These albedo patterns tend to increase $\Delta m$, but generally by < 0.1 mag.

Spencer et al. (2020) report that the distribution of normal reflectances on Arrokoth has a mean of 0.15 and a standard deviation of ~ 0.025, which is somewhat smaller than the factor-of-two variations we have simulated. The first geologic map (Fig. 1C of Spencer et al.) identifies ~ 10 large geologic units on each lobe, comparable to the number of visible triangles in the inset images in Fig. 5a. Thus, if Arrokoth is typical of small KBOs, albedo variations will generally only affect KBO light curves at the level of a few hundredths of a magnitude. Degewij et al. (1979) reached a similar conclusion about main belt asteroids, finding that few have color or albedo variations that contribute more than 0.03 mag to the light curve. This is important because procedures to derive asteroid shapes via light curve inversions (Ďurech et al., 2010; Ostro et al., 1988; Ostro and Connelly, 1984) assume that the albedo is uniform. Although the light curve of an individual KBO could be dominated by, for example, a single bright spot, we have no evidence to suggest that this is a common occurrence among small, geologically inactive KBOs; we have discounted such a possibility for the purposes of this broad, statistical analysis.

### *2.4 Effects of Limb Darkening*

As discussed above, the Lommel-Seeliger (L-S) law, sometimes referred to as the “Lunar law”, is quite successful in describing disk-resolved variations in surface brightness of many small bodies in the Solar System. It is used explicitly in asteroid shape inversions (Ďurech et al., 2010; Ostro and Connelly, 1984; Ostro et al., 1988). It is also the scattering law that Spencer et al. (2020) applied to Arrokoth.

As noted in Section 2.1, one shortcoming of the L-S law is that it does not exhibit limb darkening at zero phase. It is worth exploring the consequences of this particular assumption. Icy satellites, for example, often do show some limb darkening, although not so extreme as that described by a Lambert law. Veverka et al. (1986, see also Shepard, 2017) successfully describe the scattering law for bodies such as these using a “Lunar-Lambert” model, comprising a linear superposition of the L-S and Lambert laws. They note that the Lambert component only becomes important for bodies with albedo > 0.5, and it never dominates the L-S component except for Enceladus, which has an unusually high albedo.

For our purposes, we can treat the Lambert law as a bounding case—one that exhibits more limb darkening than we are likely to encounter in the Kuiper Belt. Figure 6 compares model light curves for equal-sized contact spheres and equivalent ellipsoids using the two laws. We find that changes in $\Delta m$ never exceed 0.065 magnitudes. For known icy satellites other than Enceladus, the weighting of the Lambert component of the scattering law, relative to the L-S component, never exceeds 40% (Veverka et al., 1986); this means that a more realistic upper limit on the potential implications of limb darkening in the Kuiper Belt is 40% smaller, i.e., ~ 0.026 mag.

### *2.5 Phase Angle Effects*

Due to their great distance, KBOs can only be observed from Earth at phase angles $\alpha \lesssim 2°$. We have explored the implications of phase angle variations on our models by generating additional light curves for all shape models using light sources that were offset from the line of sight by 2° upward, downward, rightward, and leftward in the sample images (insets in Figs. 2–6). Across a variety of shape models, the changes in $\Delta m$ had a root-mean-square (RMS) value of 0.009 mag; no single change exceeded 0.022 mag.

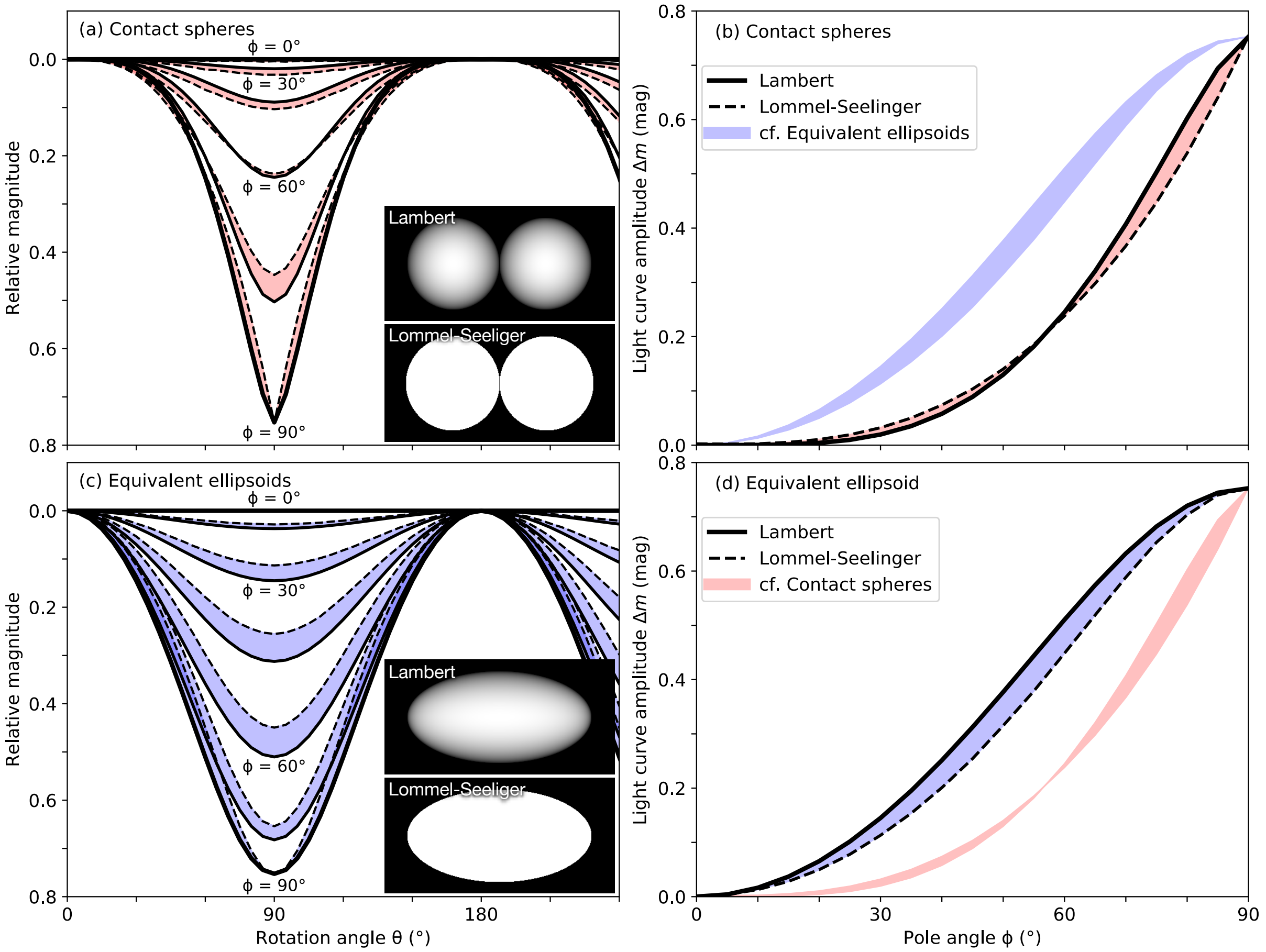

**Figure 6**. Simulations of the effects of limb darkening. We compare light curve models for bodies whose surfaces obey a Lambert law to those that obey the Lommel-Seeliger law. Our models are based on equal-sized contact spheres (panels a,b) and 2:1:1 ellipsoids (c,d). Panels on the left show rotational light curves for values of pole angle $\phi$ in steps of 15°; panels on right show the relationship between light curve amplitude $\Delta m$ and $\phi$. The shaded zone of panel d is duplicated in panel b, and vice-versa, to make it easier to compare the two sets of models.

Note that a body's phase function does not factor into this calculation. That phase function describes how the body dims overall as $\alpha$ increases, but it does not affect the amplitude of the light curve (when expressed using a logarithmic scale such as magnitudes) directly. However, the argument above is invalid if the phase function varies among the surface elements of a body. For example, if bright areas have a flat phase function slope, whereas darker areas have a steeper slope, then an increase in the phase angle would accentuate their differences. However, any such effect would have to build upon pre-existing albedo variations of the sort discussed above (Section 2.3 and Fig. 5). Belskaya et al. (2008) found phase function slopes in the range 0.05–0.20 mag/degree for a sample of KBOs. As a worst case scenario, consider bodies similar to those rendered for our study of albedo effects (Section 2.3), but with the darker areas having a 0.20 mag/degree slope and the brighter areas having a 0.05 mag/degree slope. Extrapolated to $\alpha = 2°$, this could enhance the difference between the bright and dark areas by a factor of ~ 1.3. For comparison, we previously estimated that albedo variations will generally only affect KBO light curves at the level of a few hundredths of a magnitude. Our worst case scenario would only increase that by another 30%.

In this paper, we have not investigated light curves at higher phase angles, although such results would be applicable to ongoing observations of distant KBOs by New Horizons (Porter et al., 2020; Verbiscer et al., 2019).

### *2.6 The Role of the "Neck"*

To explore the origin of the key differences between the light curves of ellipsoids and contact binaries, we have defined a continuous transformation between the two shapes. The free parameter is the radius of the "neck", the region of overlap between the two lobes. We have defined the free parameter $n$ as the dimensionless ratio of the neck's radius to the body's $b$ (intermediate) radius. In the inset panels of Figs. 2–6, this intermediate axis is oriented vertically. For our two-sphere model, $n = 0$ because the two lobes share no common volume. To increase the neck radius, we have transformed the two spheres into ellipsoids by extending each of them into the interior of the other along their common axis, while keeping the endpoints of the lobes fixed. The neck is defined as the region where the ellipsoids overlap so, at $n =$ 100%, the two bodies overlap completely, yielding a single ellipsoid. Throughout the transformation, we ensure that all models retain the same geometric cross section along all three principal axes.

The leftmost column of Fig. 7 illustrates this transformation by comparing 3-D models for various values of $n$. The additional columns of the figure show the same bodies as they are rotated toward $\theta = 90°$, where our view is aligned with the long axis. To the eye, even a body with a 90% neck generally bears a closer resemblance to touching spheres ($n$ = 0%) than it does to an equivalent ellipsoid ($n$ = 100%). Necks smaller than ~ 50% are almost indistinguishable from touching spheres.

We make this result quantitative in Fig. 8, which shows the light curve properties for transitional steps between ellipsoids and contact spheres. For a 50% neck, the increase in $\Delta m$ (Fig. 8b) is ≤ 0.036 mag; for a 70% neck, it is ≤ 0.067 mag. Fig. 8a shows that the characteristic sharp, "V"-shaped brightness minimum of the light curve persists even for a 90% neck. This is related to the fact that, in Fig. 7, a body with a 90% neck is still visibly bilobate.

The ~ 5 km diameter of Arrokoth's neck corresponds to $n \approx 50\%$. It suggests that the two lobes merged very slowly (McKinnon et al., 2020). Higher-speed mergers in the Kuiper Belt would probably result in larger necks. However, this analysis reveals that even a KBO with a very large neck can still produce light curves with distinctive broad peaks in brightness along with "V"-shaped minima, suggesting that Earth-based observers would still be able to recognized it as a contact binary.

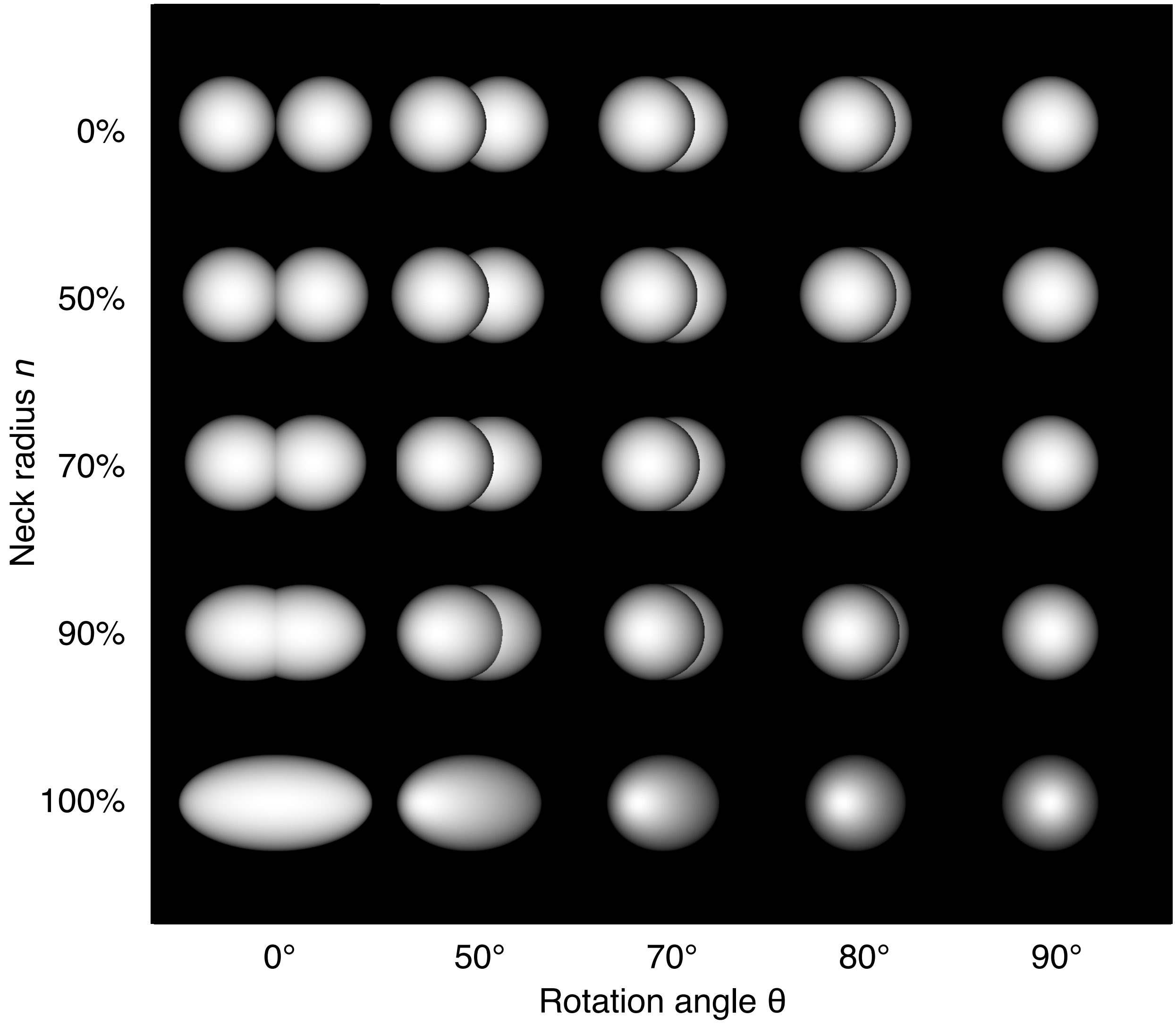

**Figure 7**. Simulated images illustrating the role of the neck. Neck radii are identified at left, where 0% defines two touching spheres and 100% defines an ellipsoid. The viewpoint is perpendicular to the rotation axis ($\phi$ = 90°). The rotation angle is indicated at bottom, where $\theta$ = 0° for a view along the short axis and $\theta$ = 90° for a view along the long axis. The figures illustrated the degree to which any neck radius ≤ 90% bears a higher degree of resemblance to touching spheres than to an ellipsoid.

The modeling discussed above does not account for Arrokoth's most notable albedo feature, its bright neck. This feature occupies just ~ 1.5% of Arrokoth's projected area in the image (Fig. 1). It has a mean normal reflectance of 0.25, compared to a body-wide mean value of 0.15 (Spencer et al., 2020; see their Fig. 1). As such, it contributes just 2.5% of Arrokoth's reflectivity. Equivalently, the bright neck would contribute only ~ 0.025 magnitudes to Arrokoth's light curve if it alternates between being visible and completely obscured as Arrokoth rotates.

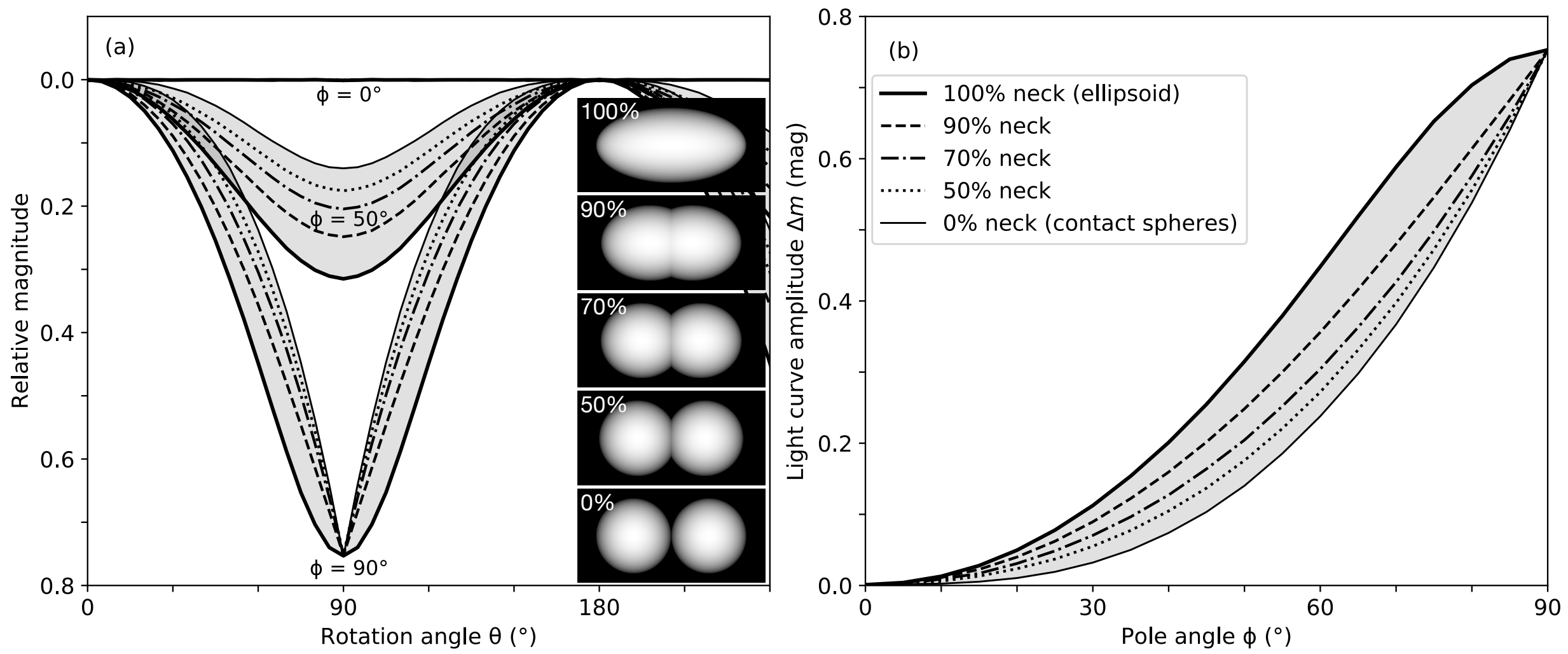

**Figure 8**. Dependence of light curve models on the radius of the neck. (a) Rotational light curves at pole angles ϕ = 0°, 50°, and 90°, for various neck radii, as identified by the dash pattern. Shaded zones identify sets of curves for the same value of ϕ. (b) Light curve amplitude as a function of ϕ. Inset images show the models for neck radii of 100% (ellipsoids), 90%, 70%, 50%, and 0% (contact spheres). Note that neck radii of 50% and 70% produce light curves that resemble those of contact spheres much more than those of ellipsoids.

## *2.7 Flattening*

Although the two lobes of Arrokoth were originally interpreted as nearly spherical, later images revealed that both are flattened, with $c/b$ = 0.5 ± 0.2 and 0.7 ± 0.2 (Spencer et al., 2020). We have explored the implications of flattening by scaling all models by factors $c/b$ = 0.8, 0.6, and 0.4. Figure 9 shows the results. Unlike the effects of topography, albedo patterns, and phase angle, flattening alters the relationship between $\Delta m$ and ϕ markedly. Although contact spheres continue to have lower light curve amplitudes overall, flattening systematically reduces $\Delta m$ at intermediate ϕ for all body shapes. Most notably, ellipsoids flattened by a factor of ~ 50% have similar light curve amplitudes to those of un-flattened contact spheres when viewed under similar circumstances.

However, it should also be noted that flattening is only an important consideration for ellipsoids with somewhat specific proportions. We have already discussed the case of prolate shapes ($b \approx c$) thoroughly. Because stable rotation must be about the short axis[2], any body with $a \approx b$ (i.e., resembling a lentil) will have a flat light curve regardless of its polar orientation and regardless of $c$. Thus, flattening is most important for bodies with three very different radii: $a \gg b \gg c$. This point is illustrated by the lowest curves of Fig. 9d, where $a/b$ and $b/c$ are ~ 2.

---

[2] Using the formulation by Harris (1994; see also Burns and Safronov, 1973; Pravec et al., 2005), the damping time for non-principal-axis rotation of an isolated, 100 km body with a 10-hour rotation period is ~ $10^4$–$10^5$ years. Damping times for larger bodies and faster rotators will be even shorter. As a result, we would not expect to find many tumbling bodies in the Kuiper Belt.

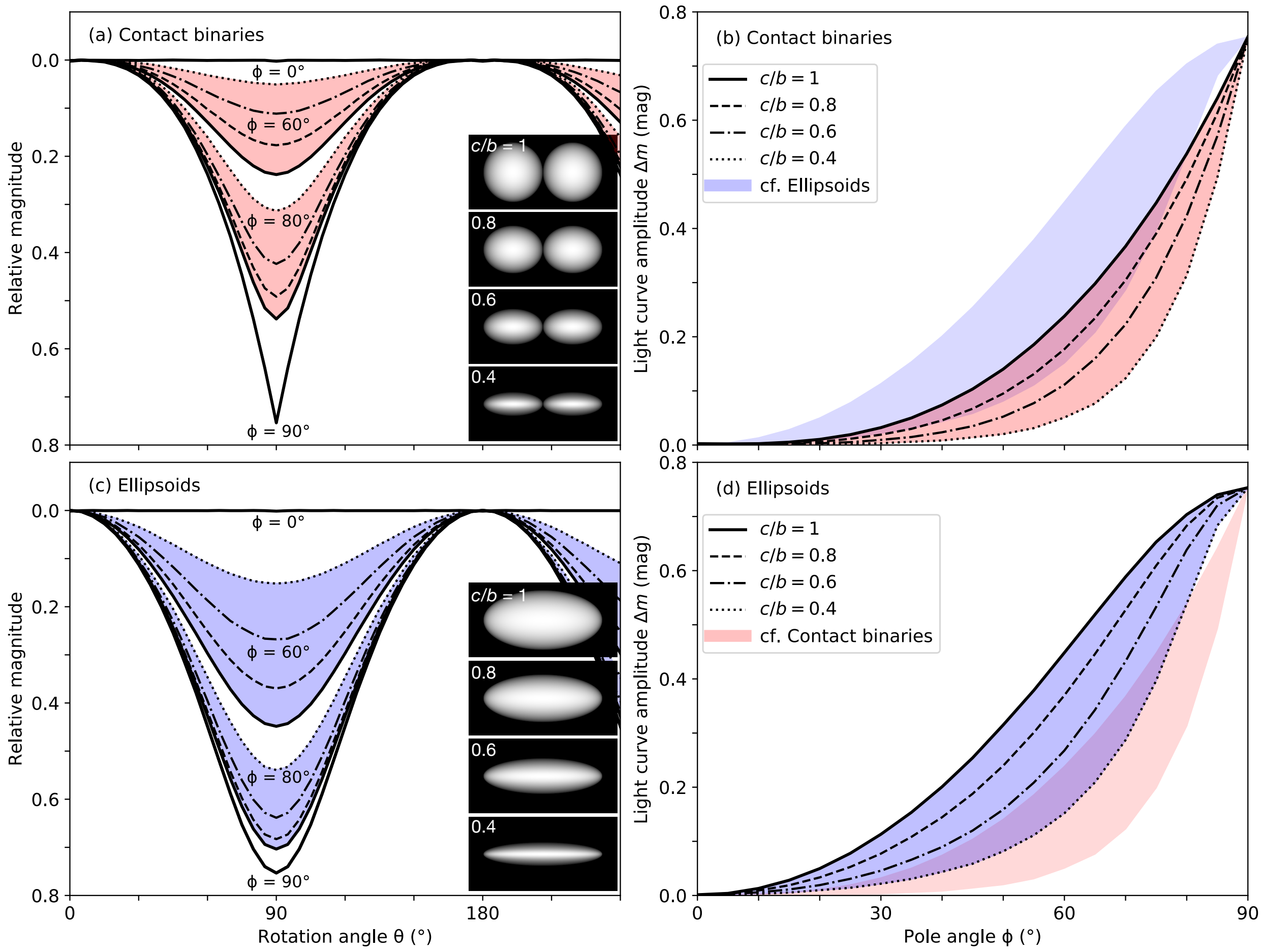

**Figure 9**. Dependence of light curve models on flattening. Left panels show rotational light curves for pole angles $\phi$ = 0°, 60°, 80°, and 90° and for four values of the ratio *c/b*. Dash patterns in the lines identify the ratio. Inset is an image of each sample body as viewed from the side, showing the longest and shortest dimensions. Shaded zones identify sets of curves using the same value of $\phi$, and illustrate the marked decrease in amplitude as the ratio decreases for intermediate values of pole angle. Right panels show the dependence of $\Delta m$ on $\phi$. The shaded zone of panel d is duplicated in panel b, and vice-versa, to make it easier to compare the two sets of models. The region of overlap indicates that flattened ellipsoids can have similar behavior to un-flattened contact binaries.

For comparison, we have investigated how common proportions like these are among the asteroids. We obtained shape models for 1620 asteroids from the DAMIT database (Ďurech et al., 2010; see https://astro.troja.mff.cuni.cz/projects/damit/) and fitted triaxial dimensions. We then selected those for which $a/b > 1.8$ to match the general circumstances of Fig. 9d (which assumes $a/b = 2$). Of the 51 asteroids satisfying this constraint, none satisfied $c/b \leq 0.4$ and only 7 (14%) satisfied $c/b \leq 0.6$. Although we do not necessarily expect KBOs and asteroids to have similar shape distributions, this test nevertheless demonstrates that ellipsoidal proportions such as these are unusual among the known small bodies in the Solar System.

### *2.8 Modeling Summary*

We have found the distinctive differences between light curves for ellipsoids and contact binaries to be surprisingly robust. Compared to ellipsoids, contact binaries consistently show broader brightness peaks and narrower minima. For radius ratios ≳ 80%, contact spheres generally show lower light curve amplitudes than ellipsoids with equivalent proportions. The quantitative effects of surface properties such as topography, albedo variations, limb darkening, and phase function generally do not exceed a few hundredths of a magnitude, at least for bodies with properties similar to those of Arrokoth.

The one important exception relates to flattening. Flattened ellipsoids can have reduced light curve amplitudes at intermediate $\phi$, making it harder to distinguish them from contact binaries. Because Arrokoth comprises two flattened ellipsoids, we must allow for the possibility that isolated, flattened ellipsoids exist in large numbers within the Kuiper Belt. We keep this particular issue in mind going forward.

## 3. Statistical Properties of Light Curves

In the previous section, we related key properties of light curves to the shapes and pole orientations of KBOs. For those relatively few KBOs with well-sampled, high-SNR, high-amplitude light curves, models such as these have already been used to infer constraints on shapes and other properties.

But can we infer anything from low-amplitude light curves? As we have already argued, it is incorrect to assume that a body is "round" simply because photometric variations are absent. If its rotation pole is pointed toward or away from Earth, a body of arbitrary shape will look the same to astronomers. Furthermore, this constraint on the pole is not especially strict: as our models (Figs. 2–9) and our experience with Arrokoth attest, an extremely elongated body, rotating about a pole ~ 40° off the line of sight (Spencer et al., 2020), can still have a very flat light curve. Thus, the only conclusion one can draw from flat light curves is probabilistic.

Because of the small number of light curves with appropriate measurements for detailed modeling, we now focus on the statistical properties of light curves produced by an ensemble of KBOs, seeking to use this information to infer general information about the population as a whole. Typically, the most readily measurable quantity is amplitude $\Delta m$. Note that, with suitable sampling, $\Delta m$ can often be determined to reasonable precision even for KBOs whose rotation periods are not yet known; the literature contains many such measurements, as we will discuss below in Section 4.2.

We represent the distribution of $\Delta m$ by a cumulative probability function $M$, where $M(x)$ is the fraction of bodies with $\Delta m \leq x$. $M(0) = 0$ and $M \to 1$ for large x. If a family of KBOs has identical physical properties but randomly oriented poles, then plots of $\Delta m$ vs. $\phi$ (right panels in Figs. 3–6, 8, and 9) can be transformed into cumulative probabilities. First, for notational convenience, we define $\phi' = \min(\phi, 180° - \phi)$. This is useful because $\phi$ can vary from 0° to 180°, but $\phi > 90°$ simply refers to retrograde spin about the supplementary pole angle 180° - $\phi$. Second, we invert the plots into functions $\phi' = f(\Delta m)$, which is possible because all curves are now monotonic. Then we can write:

$$M(x) \equiv \text{Probability}[\Delta m \leq x] = \text{Probability}[\phi' \leq f(x)] \text{ ,} \tag{1}$$

thereby relating $M$(x) to the cumulative probability of $\phi'$.

The simplest assumption one might make about KBO pole directions is that they are isotropically distributed over 4π steradians. Such a distribution has the property that cos $\phi$ is uniformly distributed between -1 and 1; this is a corollary of Archimedes' "Hat Box" problem. Restricting consideration to $\phi' \leq 90°$, the relationship is:

$$M(x) = \text{Probability}[\cos \phi' \geq \cos(f(x))] = 1 - \cos(f(x)) \text{ ,} \tag{2}$$

or, in simpler notation,

$$M(\Delta m) = 1 - \cos \phi'(\Delta m) \text{ .} \tag{3}$$

We cannot, however, be certain that KBO poles are distributed isotropically. For example, Nesvorný et al. (2019) predicted that cold classical KBOs should be preferentially prograde. To handle a constraint such as this, we define the rectangular $(x,y,z)$ coordinate frame centered on the body, with the orbit in the $(x,y)$ plane. In this frame, we define a pole vector $\mathbf{P} = (\sin \eta \sin i, -\cos \eta \sin i, \cos i)$. Here $i$ is the obliquity angle, measured from the orbit plane's normal vector to the pole: $i < 90°$ for prograde spin and $i > 90°$ for retrograde spin. The second angle, η, defines the ascending node of the KBO's equator within the orbit plane. Because Earth falls very close to the Sun on the scale of a KBO's orbit, we can safely assume that our line of sight falls within the orbit plane. For a KBO at orbital longitude ζ, its pole angle $\phi$ satisfies $\cos \phi = \mathbf{P} \cdot (\cos \zeta, \sin \zeta, 0) = \sin i \sin (\eta - \zeta)$.

Now we expand our consideration from an individual KBO to an entire population. In this case, we generally expect longitudes ζ to be uniformly distributed between 0° and 360° [3]. If so, it follows that the quantity (η - ζ) mod 360° must be uniformly distributed as well, even if η is not. For convenience, we substitute $u = 90° - (\eta - \zeta)$, so $\cos u = \sin(\eta - \zeta)$. Following (1):

$$M(x) = \text{Probability}[|\sin i|\, |\cos u| \geq \cos(f(x))] = \text{Probability}[|\cos u| \geq \cos(f(x)) / |\sin i|] \text{ .} \tag{4}$$

Because all four quadrants of angle $u$ are now equivalent, we can restrict our consideration to a uniform distribution of $u$ within the domain [0,90°]. Eq. (4) divides into two possibilities. If the rightmost term, $\cos(f(x)) / |\sin i|$, is > 1, then the probability is zero. Otherwise,

$$M(x) = \text{Probability}[u \leq \cos^{-1}(\cos(f(x)) / |\sin i|)] = 2/\pi \cos^{-1}(\cos(f(x)) / |\sin i|) \text{ .} \tag{5}$$

Reverting to simpler notation,

---

[3] As noted by Chiang and Jordan (2002), resonant KBOs are not uniformly distributed in ζ due to their nonzero eccentricities and their interactions with Neptune. Such bodies might require more careful modeling than we provide here.

$$M(\Delta m) = 0 \qquad \text{for } \phi'(\Delta m) < 90° - i\,; \tag{6a}$$

$$= 2/\pi \cos^{-1}(\cos \phi'(\Delta m) / |\sin i|) \qquad \text{otherwise.} \tag{6b}$$

A few brief comments about Eqs. (6) are in order. The special case where $i = 90°$ describes a Uranus-like pole that lies “sideways” in the orbit plane. In this case, (6a) is never used and (6b) reduces to:

$$M(\Delta m) = 2/\pi\ \phi'(\Delta m)\ . \tag{7}$$

This linear relationship between $M$ and $\phi'$ simply states that the distribution of $\Delta m$ derives from the fact that all values of $u$ are equally likely, and therefore all values of $\phi$ are equally likely. At the other extreme where $i = 0°$, the pole is fixed and perpendicular to the line of sight, meaning that $\phi' = 90°$ is the only value. Here (6a) says that $M = 0$ for $\phi' < 90°$, but (6b) says that $M = 1$ for $\phi' = 90°$. This discontinuous function is exactly what one would expect for a cumulative probability function describing a fixed value.

More commonly, obliquity $i$ would be described by a probability density function (PDF) rather than taking on a single, known value. Suppose the distribution of $i$ is defined by a PDF $h(i)$, meaning that the likelihood of finding an obliquity between $i$ and $i + di$ is $h(i)\ di$. The full expression for $M(\Delta m)$ is then a weighted sum over the formulas in Eq. (6), using $h$ as the weight. This formula can be readily evaluated numerically.

Figure 10 illustrates the implications of these formulas for the models shown in Fig. 3 above. For a particular $\Delta m$ on the horizontal axis, each curve identifies the fraction of KBOs expected to have amplitudes smaller than this value. In effect, we have taken the curves of Fig. 3b, flipped them about the diagonal, and then transformed the new vertical axis based on formulas presented above. For the isotropic assumption (Fig. 10a), the vertical axis is equal to $1 - \cos \phi'$ (Eq. 3; see right axis in Fig. 10a). Figure 10b shows the same shape models but for an alternative distribution of poles in which $i$ is uniformly distributed between 0 and 180° (or, equivalently, $h(i) = 1/\pi$ for $i$ in $[-\pi/2,\pi/2]$). This distribution increases the number of bodies with $i$ near 0° and 180°, for the same reason that longitude lines get closer together near the poles on a globe of the Earth. This, in turn, increases the likelihood that bodies are observed from a viewpoint nearly perpendicular to their rotation poles, resulting in systematically higher values of $\Delta m$ in Fig. 3b than in Fig. 3a.

## 4. Comparisons to the Available Data

### *4.1 Pole Distributions*

Aside from Arrokoth, only two single KBOs have constrained rotation poles. Lacerda (2011) noted a change in the light curve of Plutino (139775) 2001 $QG_{298}$ and determined that this could be explained by Earth’s changing viewpoint on the body’s pole if obliquity $i = 90° \pm 30°$. Using similar methods, Fernández-Valenzuela et al. (2019) derived an obliquity of ~ 150–160° for (20000) Varuna. For comparison, Arrokoth also has a high obliquity of 99° (Spencer et al., 2020). Thus, although we currently have very little information about the directional distribution of KBO poles, we can see that the first three measurements are widely distributed.

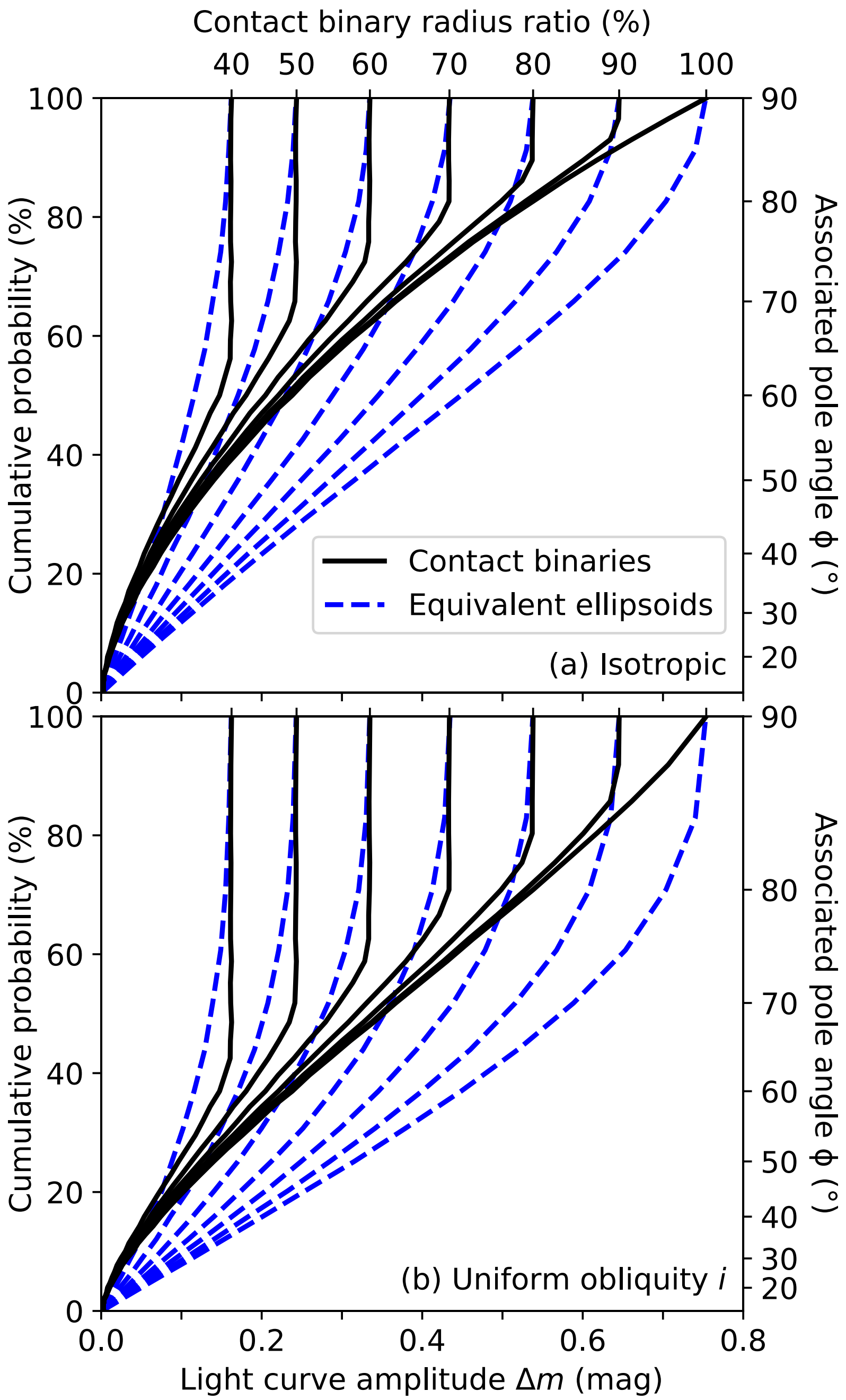

**Figure 10**. Cumulative probability functions, i.e., the fraction of KBOs expected to have light curve amplitudes lower than the specified $\Delta m$, depending on a shape model and pole distribution. Shape models are as presented in Fig. 3: solid for contact binaries and dashed for equivalent ellipsoids. Curves are shown for (a) an isotropic distribution of poles and (b) poles with uniformly-distributed obliquities $i$. The right axis shows the pole angles that are associated with the cumulative probabilities along the left.

In addition to the above, the pole of one Centaur, (10199) Chariklo, has been determined based on observations of its ring system (Leiva et al. 2017); its obliquity is ~ 40°. However, because Centaurs have orbital histories that probably include at least one close encounter with a planet (Gomes et al., 2008; Peixinho et al., 2020), we cannot be sure that Chariklo's orientation is primordial.

An alternative guess about how KBO poles are distributed is to suppose that they have a similar directional distribution to the orbital angular momentum vectors of multiple KBOs. Grundy et al. (2019) have recently summarized the orbits of 35 binaries. One key finding is that orbits are far more likely to be prograde than retrograde. This is consistent with the dynamical models by Nesvorný et al. (2019), which predict that binary cold classicals should have preferentially prograde orbits.

In their analysis, Grundy et al. (2019) found it useful to distinguish between "tight" and "wide" binaries, distinguished by the separation distance of each pair. For the purposes of this analysis, we focus on the tight binaries, presuming that they serve as a better analogue for the rotation states of single KBOs. Grundy et al. find that the distribution of angular momentum poles of tight binaries is reasonably well matched by what might be described as a "half-isotropic" distribution. Specifically, if they replace the inclinations *i* of all retrograde orbits by 180° - *i*, then the resulting distribution is roughly uniform over 2π steradians. Because our analysis cannot distinguish between prograde and retrograde rotations, this is equivalent to saying that an isotropic distribution is a reasonable approximation for our purposes. However, our alternative distribution, assuming a uniform distribution of inclinations (Fig. 10b) is also generally compatible with Grundy et al.'s statistical sample; see their Fig. 3, where the uniform distribution would appear as a straight diagonal.

### *4.2 Light Curve Amplitudes*

We now apply our analytical framework to the current literature on KBO light curves. Table 1 shows the assembled information. For each KBO or Centaur, we list the absolute magnitude and the measured light curve amplitude $\Delta m$ along with its quoted uncertainty. This summary encompasses results from several recent surveys, plus additional numbers found in the literature; citations are identified in column "Ref". (Note that these refer to the paper from which we obtained the numbers shown, not necessarily to the original observations.) Upper limits are indicated by a less-than sign. Some published measurements are lower limits, indicated by a greater-than sign. For internal consistency, we have derived all of our classifications using the criteria established by Elliot et al. (2005) for the Deep Ecliptic Survey[4]. Three bodies do not have well determined classifications and are identified in the table as unknown. We include Centaurs in our tabulation because these bodies are probably recent escapees from the Kuiper Belt (Gomes et al., 2008; Peixinho et al., 2020).

We have assigned each available measurement a weight of either 0 (rejected) or 1 (accepted). We have rejected measurements based on reasonably objective criteria as follows. (1) Because we are interested in the shapes of individual bodies, we ignore measurements of known binaries, obtained from a list that is maintained on line by W. M. Grundy of Lowell Observatory (http://www2.lowell.edu/users/grundy/tnbs/status.html); see Noll et al. (2020) for details. (2) Larger bodies may be deformed by their own self-gravity, so we reject those with absolute magnitude $H < 5.5$, corresponding to radius of a few hundred km (for albedo ~ 0.1). We would prefer to set a lower size threshold (higher $H$), but this is a limit that still provides us with reasonable statistics. (3) For bodies with multiple light curve measurements in the literature, the table contains the most recent and/or most precise measurement as long as prior measurements are roughly compatible. We have rejected bodies for which measurements in the literature are mutually inconsistent. (4) In a similar vein, we reject measurements of bodies whose light curves appear to be variable. (5) Some amplitudes are identified as a lower limit. This generally indicates that the body shows distinct variations but that no periodicity has yet been identified. While it is true that the existing observations may have missed a large change, this becomes less likely as the number of observations increases. Because typical KBO rotation periods are less than 12 hours (Thirouin et al., 2016; Thirouin and Sheppard, 2018, 2019), and most light curves are double-peaked, the time interval between light curve extrema is typically only ≲ 3 hours. Here, we have rejected lower-limit measurements for which all observations occurred on a single night. However, when the observations span two or more nights, we include the body in our statistical analysis even though the value of $\Delta m$ may be somewhat underestimated. Using similar reasoning, Alexandersen et al. (2019) argued that most of their two-night observations captured 80–88% of the full light curve amplitude. In the table, the note column indicates our reasoning behind each judgment call.

[4] Classifications for individual objects, which are updated with new astrometric releases by the Minor Planet Center, can be found online at https://www.boulder.swri.edu/~buie/kbo/desclass.html. The classifications used in Table 1 are up to date as of April 2020.

Figure 11a shows the cumulative amplitude distributions derived from Table 1. The $\Delta m$ for each KBO is represented by a minimum and maximum value (always in magnitudes) as listed in the table. For measurements given as upper limits, we assume that the lower limit is zero. If error bars are provided, they define the minimum and maximum. Other values are treated as exact. When converted to a cumulative distribution, each KBO is represented by a linear ramp from the minimum to the maximum value, and the curves are just the normalized sum of all these ramps. Bodies without error bars create vertical jumps in the plot, but these jumps are otherwise harmless.

In the table and plot, bodies are classified as classicals, resonant bodies, scattered disk bodies, and Centaurs. The first three categories have almost identical distributions, so we also show them combined. Centaurs seem to be a bit different, with a larger fraction having lower amplitudes. We have included the Centaurs in this study primarily for completeness; because their source region is widely believed to be the Transneptunian region (Gomes et al., 2008; Peixinho et al., 2020), comparing their overall properties with those of their potential source region may prove useful in future analyses.

In this analysis, however, we have opted to plot the measurements from the Outer Solar System Origins Survey (OSSOS) separately from the others (Alexandersen et al., 2019). These show a very different cumulative distribution, with a substantially larger fraction having high amplitudes. For example, in the other measurements we have assembled, which correspond to the solid blue curves in Fig. 11, more than 50% of KBOs have $\Delta m < 0.2$ mag, whereas only 10% are this low in the OSSOS survey. No measurement in the survey is closer to zero than five

**Figure 11**. The cumulative distribution of KBO light curve amplitudes and its comparison with various models. Panel a shows observed cumulative distributions as derived for this analysis, based on 20 classicals, 23 resonant bodies, 13 scattered bodies, 36 Centaurs, and 60 OSSOS KBOs, all identified in Table 1. The "combined" curve includes everything in the first three categories, plus the three additional unclassified KBOs in Table 1, for a total of 59. We plot the OSSOS data set separately for reasons discussed in Section 4.2. Panels b–i compare three observed distributions to models involving contact binaries of various radius ratios (left panels) and ellipsoids with various proportions (right panels). Panels b–e assume an isotropic distribution of rotation poles, whereas f–i assume that obliquities *i* are uniformly distributed. Rows alternate between models for un-flattened shapes and those flattened by *c*/*b* = 0.6.

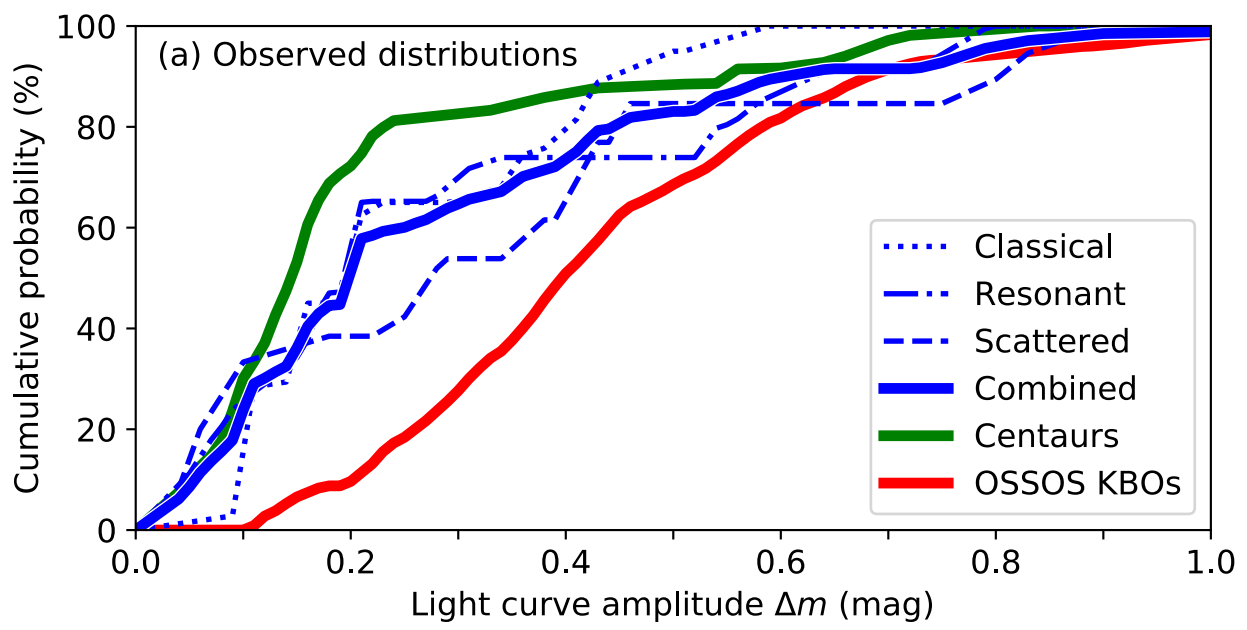

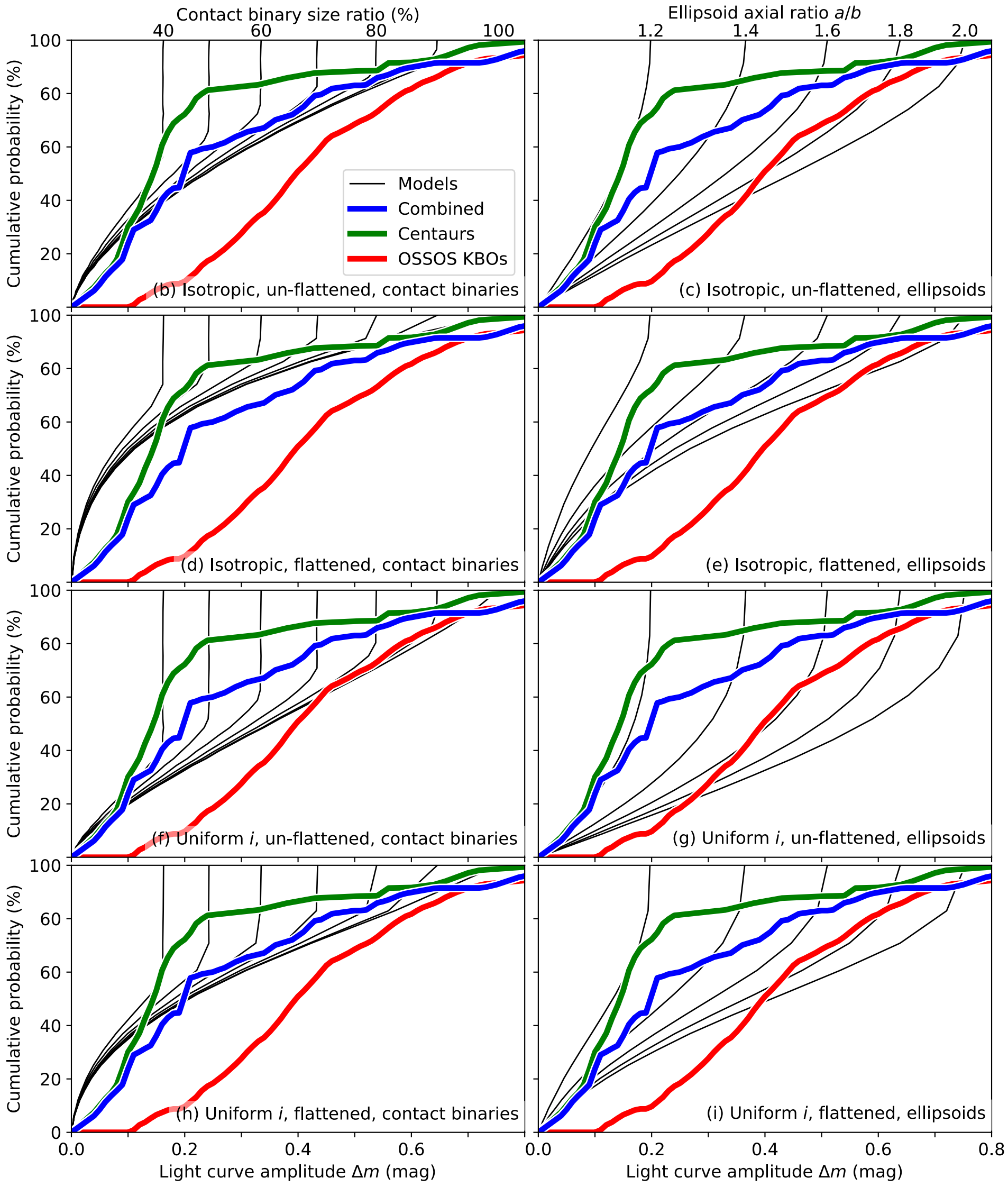

times the measurement uncertainty.

We would have expected greater consistency between the OSSOS and non-OSSOS measurements. The Kolmogorov-Smirnov (K-S) statistic can be used to test the hypothesis that two sets of samples have been drawn from the same probability distribution (Hodges, 1958). We have performed this test on the $\Delta m$ values for the combined classical, resonant, and scattered KBOs in each data set. Where individual values of $\Delta m$ are uncertain, we have used the mean. Using the implementation of the K-S statistic provided in Python module scipy.stats.ks_2samp, we can reject the hypothesis that the OSSOS and non-OSSOS measurements are samples of the same distribution with 99.9997% confidence ($p$-value = $3 \times 10^{-6}$).

Part of the explanation for the difference may be that Alexandersen et al. (2019) calculated $\Delta m$ from the difference between the highest and the lowest measurements; this increases the likelihood that noise or one spurious measurement could artificially increase $\Delta m$. However, this is not the entire explanation. Thirouin and Sheppard (2019) made their own measurements of the cold classicals in the OSSOS data set and, while some of their amplitude determinations are lower, it is still true that only one of 25 light curves has $\Delta m < 0.2$ mag. Using the K-S test, we can have 99.4% confidence that OSSOS measurements by Thirouin and Sheppard's do not sample the same probability distribution as the non-OSSOS classicals in Table 1.

### *4.3 Statistical Comparisons*

In order to proceed, we need an unbiased sample of KBOs—one in which neither high-amplitude nor low-amplitude light curves have been favored. There are many reasons why this sample might contain biases. (1) Because it takes multiple detections to confirm a discovery, a faint KBO with high $\Delta m$ may not be consistently detectable, whereas the same body with lesser variations would be. (2) There may be a human bias in favor of reporting highly variable KBOs, simply because their light curves are more interesting and informative. (3) Although we have filtered out known binaries from this analysis, our sample could still contain unresolved binaries. The prevalence of unresolved binary KBOs is currently unknown. Noll et al. (2014) suggested a very high binary fraction for $H \leq 6.1$ mag, but only ~ 20% for smaller KBOs. More recently, early results from a survey of 200 cold classical KBOs (Parker et al. 2019) does not support such a large binary fraction (Benecchi, S., personal communication, 2020). The effect of this bias on our statistics is unknown, because the literature contains examples of binaries with both large and small light curve amplitudes (Thirouin et al., 2014; see their Table A.2). For this initial study of KBO light curve statistics, we proceed on the assumption that our sample is unbiased simply because we have no clear alternative. Future, larger-scale surveys of KBOs should provide a more reliable body of data for the type of analysis presented here.

In Figs. 11b–i, we compare the measured distributions of light curve amplitudes with the models that we have developed. Our three measured, cumulative distributions (combined, Centaur, and OSSOS of Fig. 11a) are shown in each panel, compared to eight sets of models: contact binaries vs. ellipsoids, un-flattened shapes vs. flattened shapes (using $c/b = 0.6$), and obliquity distributions that are either isotropic or uniform. Each black curve shows a model in which every KBO has exactly the same shape. The curves are organized in families, where individual curves are defined by the ratio of the lobe radii of contact binaries (40–100%) or the axial ratio *a/b* of ellipsoids (1.2–2).

A KBO population comprising a diverse set of shapes could always be represented by linear combinations of curves such as these. It follows, however, that a family of shape models is not compatible with any measured distribution that falls below its lowest curve. As a result, the OSSOS data set is incompatible with any of the models we have presented. The models of panel g (un-flattened ellipsoids with a uniform distribution of obliquities) come closest to being compatible, but even then only if we also assume that the OSSOS data set under-samples KBOs with $\Delta m \lesssim 0.2$ magnitudes. An alternative way to state the result is this: the OSSOS photometry implies that KBOs have a strong preference for low obliquities and axial ratios $a/b \gtrsim 2$.

For comparison, the blue line in Fig. 11 represents measurements of other classicals, resonant bodies, and scattered disk KBOs combined. Among the eight panels b–i, it is notable that the blue curve shows a striking match for the lowest curve in panel b, meaning that the observed distribution of KBO light curves is consistent with what one would expect for a population of contact binaries with spherical lobes of equal size, having an isotropic distribution of poles. However, panels d and h show that a population of flattened contact binaries, similar to Arrokoth, is not compatible with the data. In each of the remaining five panels, it would be possible to construct a mixture of family members with different shape ratios to match the cumulative distribution, so no solution is unique and it is not possible to draw definitive conclusions.

Centaurs, shown in green in Fig. 11, could be successfully described by any of these families of models. What is most notable about Centaurs is the simple fact that they are noticeably different from KBOs, with a higher fraction of low-amplitude light curves. According to the K-S test, the hypothesis that these two data sets sample the same distribution of $\Delta m$ can be rejected with 94% confidence ($p$-value = 0.06), a difference that is strongly suggestive if not entirely conclusive. Perhaps the coma-like activity often observed around Centaurs tends to reduce or obscure their photometric variations. However, further analysis and interpretation is beyond the scope of this initial investigation.

### *4.4 Amplitude and Absolute Magnitude*

The light curve models discussed in Section 2 have another important implication. Because a body's stable rotation state is always about its shortest axis, it generally shows its largest cross section when observed by looking down the rotation pole ($\phi$ = 0° or 180°). From this viewpoint, the projected area $A$ is $\pi ab$ for an ellipsoid and $\pi(r_1^2 + r_2^2)$ for a contact binary, where $r_1$ and $r_2$ are the radii of the two lobes. These are the maximum possible values of $A$, making each body as bright as it can be in our photometric models. Flattening would not modify either value, because the flattened axis is the rotation axis, which is parallel to the line of sight.

Because $A$ is maximized and also independent of the rotation angle θ, we can state this is a succinct rule: *a given body tends to be brightest when its light curve is flattest*. The flat light curves labeled by $\phi = 0$ in Figs. 2, 4, 5, 6, 8, and 9 all illustrate this rule.

Consider now what happens when we change $\phi$ to 90°. From this viewpoint, an ellipsoid will alternate between $A = \pi ac$ and $\pi bc$, thereby maximizing its light curve amplitude at a value $\Delta m = 2.5 \log_{10}(a/b)$ mag (Fig. 2b). This viewpoint provides the observer with the body's smallest possible time-averaged $A$, because its shortest axis, $c$, is always visible, and its longest axis, $a$, is oriented along the line of sight for part of each rotation period. Similarly, an un-flattened, spinning contact binary will alternate between $A = \pi(r_1^2 + r_2^2)$ and $A = \pi r_1^2$ (assuming $r_1 \geq r_2$), for a maximized amplitude $\Delta m = 2.5 \log_{10}(1 + r_2^2/r_1^2)$ mag (Fig. 2a). Time-averaged $A$ is again minimized, because this point of view maximizes the fraction of time during which the larger lobe obscures the smaller. If the binary is flattened, $A$ will be reduced even further, but $\Delta m$ will be unchanged because it only depends on the ratio $r_2/r_1$. In brief, *a given body's light curve amplitude tends to be maximized when its mean brightness is minimized, and vice versa*. Note that the light curves in Figs. 2, 4, 5, 6, 8, and 9 all have not just their largest variations, but also their lowest mean brightnesses, at $\phi = 90°$.

These two simple rules, which define a strong anticorrelation between the time-averaged brightness of a given body and its light curve amplitude, have important implications. We proceed for now by restricting our consideration to un-flattened bodies; we will discuss the implications of flattening in the next section, 4.5. Our un-flattened models (Figs. 2, 4, 5, 6, and 8) all have the property that a body's maximum brightness at $\phi = 90°$ is roughly equal to its fixed value at $\phi = 0°$. This is consistent with the arguments above based on cross-sectional areas. For example, if $b = c$, then an ellipsoid's maximum $A$ is always $\pi ab$, regardless of $\phi$. Similarly, the maximum $A$ for un-flattened, contact spheres is always $\pi(r_1^2 + r_2^2)$. It follows that *the most stable photometric attribute of a given body is the brightest point on its light curve*. This corresponds to the body's minimum in absolute magnitude, $H_{min}$. The relationship between $\Delta m$ and the time-averaged absolute magnitude, $H_0$, can be written simply as $H_0 \approx H_{min} + \Delta m/2$; note that, because of the negative sign in the definition, $H_0$ and $\Delta m$ actually have a positive correlation.

The literature contains several reports of observed correlations between absolute magnitude and $\Delta m$ (Alexandersen et al., 2019; Benecchi and Sheppard, 2013; Thirouin and Sheppard, 2019), which have been interpreted as evidence that smaller KBOs are generally more irregular in shape than large ones. The implied assumption here is that $H_0$, the time-averaged value, can be used as a rough proxy for size. We have shown that the minimum value, $H_{min} = H_0 - \Delta m/2$, is a much more reliable proxy, because it is not as dependent on the orientation of each body's pole. The first requirement for any size metric should be to identify two identical bodies as having the same size. For un-flattened bodies, $H_{min}$ meets this goal; $H_0$ does not. For more diverse shapes, all size metrics are imperfect, but $H_{min}$ at least partially compensates for pole orientation effects, which are a known source of error. Furthermore, we know that $H_0$ and $\Delta m$ are directly coupled—$H_0$ goes up when $\Delta m$ goes up—and this bias could be contributing to the correlation that others have noted.

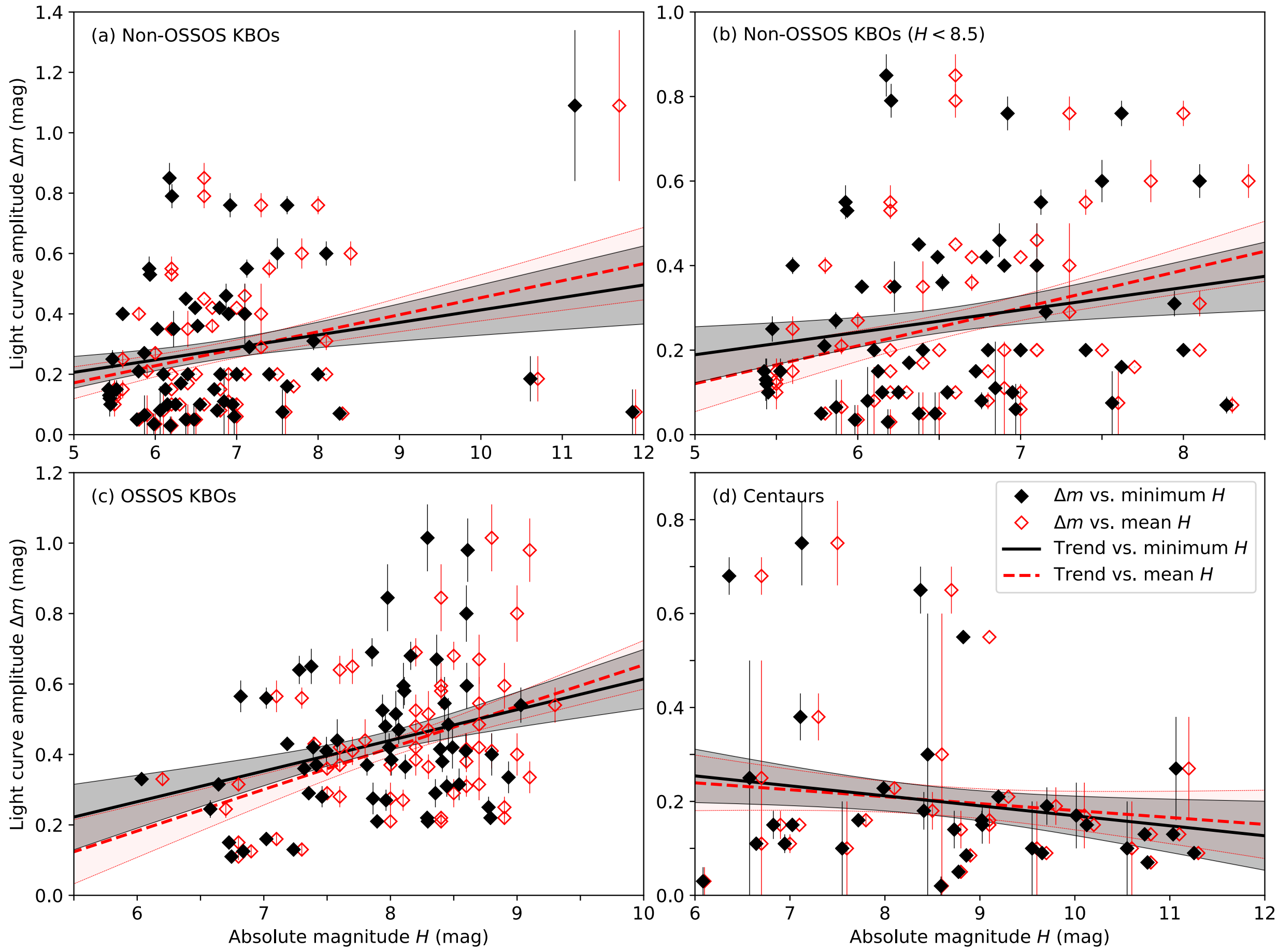

**Figure 12**. Scatter plots of light curve amplitudes $\Delta m$ versus the mean (open symbols) and minimum (filled symbols) values of absolute magnitude $H$, as tabulated in Table 1. As we argue in the text, minimum $H$ is the preferred proxy for KBO size. Dashed lines are linear fits to the open symbols; solid lines are linear fits to the filled symbols. The uncertainty (± 1σ) in each linear fit is indicated by the shaded zone centered on each line. The solid-line fits have consistently lower slopes than the dashed-line fits, illustrating the degree to which reported correlations between $\Delta m$ and KBO size might be artifacts of the earlier analysis methods.

We revisit this question in Fig. 12, which contains scatter plots of $\Delta m$ versus $H_0$ and $H_{min}$ from Table 1. For each alternative abscissa, we fit a straight line to all of the measurements and assess the slope of the trend. In the plots, the shaded region around each trend line shows the one-sigma uncertainties in that fit, based on the covariance matrix[5]. A positive slope, significantly different from zero, would support the hypothesis that the trend is real. Our null hypothesis is that there is no correlation between size and $\Delta m$.

[5] For a linear regression fit $y = ax + b$, we can determine the variance as $\mathrm{var}(y) = \mathrm{var}(a)\, x^2 + \mathrm{cov}(a,b)\, x + \mathrm{var}(b)$. The three coefficients in this quadratic function are elements of the covariance matrix.

Panel 12a shows the non-OSSOS KBOs, corresponding to the blue curves in Fig. 11. The trend-line slopes are 0.056 ± 0.024 when using $H_0$ as the abscissa, but 0.041 ± 0.025 when corrected to $H_{min}$. Although the first slope has 2.5σ significance ($p$-value = 0.007), the second slope is the better test of whether $\Delta m$ is actually correlated with size; with significance of 1.7σ and a $p$-value of 0.04, this result does not give us as firm a basis to reject the null hypothesis.

In the figure, each measurement appears twice: once as an open symbol at coordinates ($H_0$, $\Delta m$) and again as a closed symbol at ($H_0 - \Delta m/2$, $\Delta m$). Of course, individual values of $H_0$ are uncertain; formal error bars are rarely quoted in the literature, but discrepancies at the level of up to 0.3 magnitudes are commonplace. Nevertheless, because the separation between each pair of points in Fig. 12 is always $\Delta m/2$, errors in $H_0$ will only produce small, pairwise, horizontal shifts of the points. Such errors could potentially alter the precise slopes and $p$-values identified above, but the marked difference between the two inferred trends will remain.

The fits in Fig. 12a are dominated by the long "moment arm" exerted by the three rightmost measurements, which refer to unclassified KBOs found in a very deep survey using the Hubble Space Telescope (Trilling and Bernstein, 2006). One of these, 2003 $BF_{91}$ (upper right in Fig. 12a), has the highest light curve amplitude in our data set. Figure 12b shows the result of omitting these measurements. The slope of the corrected ($H_{min}$) trend line is 0.053 ± 0.039 ($p$-value = 0.09), at best only marginally significant. We also conducted the same analysis after omitting those KBOs with a "greater-than" constraint in Table 1, in order to assess the possibility that underestimated values of $\Delta m$ may bias our results. Those results are similar, producing a fitted slope of 0.066 ± 0.044 ($p$-value = 0.07).

The situation is different in Fig. 12c, which shows our analysis of the KBOs in the OSSOS survey (Alexandersen et al., 2019). Here the trend line is strong, with a corrected slope of 0.087 ± 0.037, for 2.3σ significance; this provides 99% confidence ($p$-value = 0.01) that this trend is real. Once again, we are led to different conclusions from the OSSOS and non-OSSOS data sets. We discuss this topic further in Section 5.1. Nevertheless, this figure still illustrates the importance of applying the correction; the slope associated with $H_0$, rather than $H_{min}$, is 0.118 ± 0.034, which would have provided a misleading indication of even higher significance, with $p$-value = $2.4 \times 10^{-4}$. If we restrict our consideration to OSSOS KBOs ≤ 8.5 mag, as we did in Panel 12b, the trend line slopes do not change very much: the slope based on $H_{min}$ becomes 0.094 ± 0.048 ($p$-value = 0.03) and that based on $H_0$ becomes 0.128 ± 0.043 ($p$-value = 0.001).

Finally, for completeness, we show our analysis of the Centaur population in Fig. 12d. Here the slope is negative, although still consistent with zero: -0.021 ± 0.019. Because Centaurs are closer to us and therefore detectable over a broader range of absolute magnitudes, any correlation between size and $\Delta m$ ought to be more easily detected for Centaurs than KBOs. The results are instead consistent with the null hypothesis.

### *4.5 Flattening, Amplitude and Absolute Magnitude*

Above, we assumed that all shapes are not flattened, so $b \approx c$. As Fig. 13 illustrates, flattening can alter the relationship between $H$ and $\Delta m$ substantially. This figure is identical to Fig. 2 except that $c/b = 0.6$ for both shapes. Here, the brightest point on each body's light curve is no longer conserved as $\phi$ approaches 90°—the maxima and the minima both decrease. At $\phi = 90°$, the overall light curve has been reduced in flux by a factor $c/b$, which amounts to 0.55 mag in this case. However, because $c/b$ just serves as an overall scale factor on the body's projected area at $\phi = 90°$, the light curve's amplitude, expressed in magnitudes, is the same as before.

We noted that the mean of each light curve in Fig. 2 is roughly $\Delta m/2$. This is because the mean falls roughly half-way between the extrema, but the brightness maxima are fixed. In the sequence of light curves in Fig. 13, we see that the maxima of the light curves decrease by 0.55 mag at the same time that ϕ increases from 0 to 90° and $\Delta m$ increases from 0 to 0.75 mag. If we express this as a relationship between the mean of the light curve and $\Delta m$, then the proportionality constant $C$ becomes 0.5 + 0.55/0.75 = 1.23. Here the first term, one half, applies to all bodies (cf. Fig. 2) and the second accounts for the additional change associated with flattening.

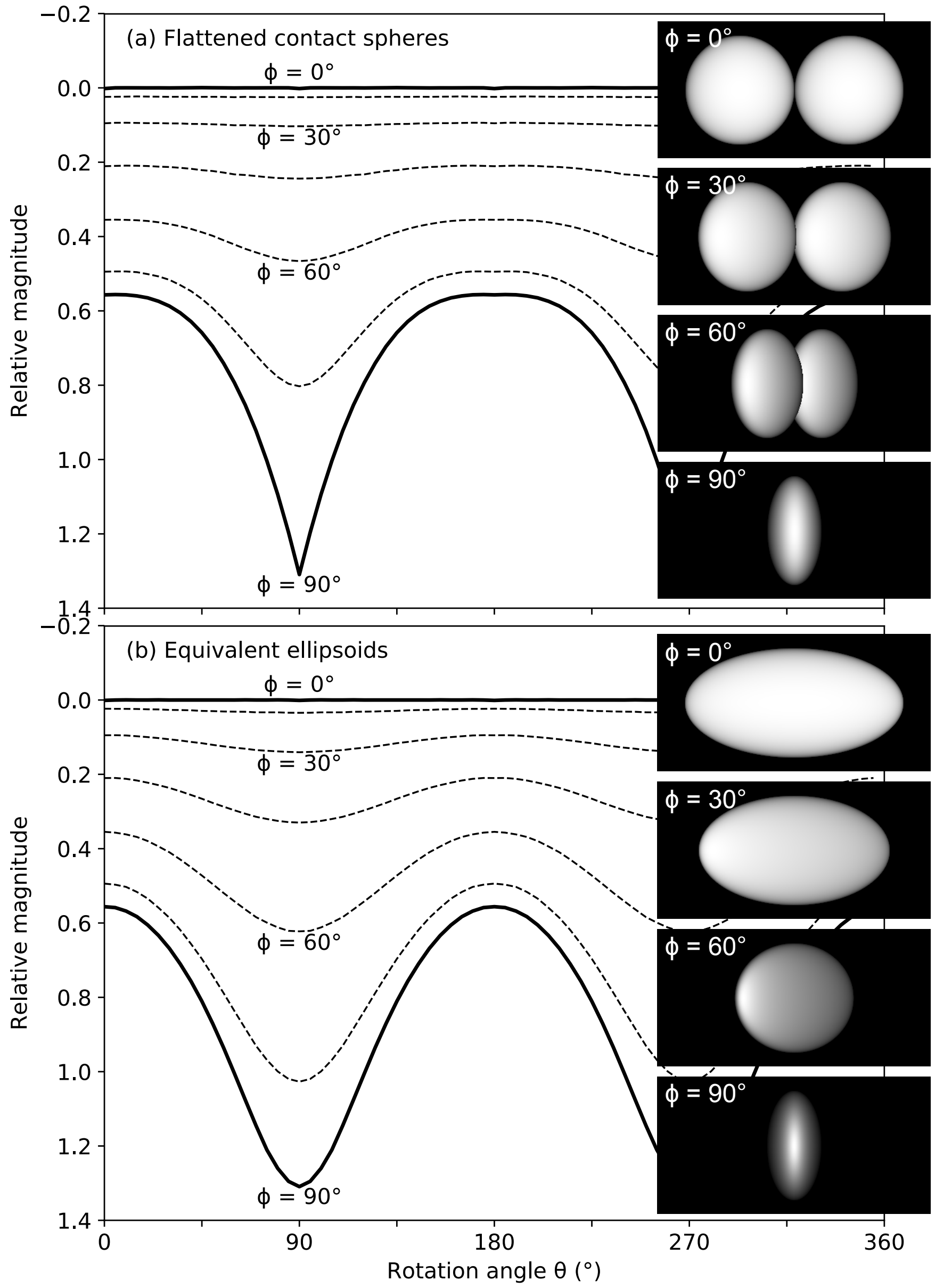

**Figure 13.** Simulated rotational light curves for two idealized shapes, each flattened by a factor $c/b = 0.6$ compared to the equivalent bodies in Fig. 2. The notable change is that the light curves no longer share their brightest points in common (their minima in magnitudes) as pole angle $\phi$ increases from 0° to 90°. Instead, bodies decrease more in overall brightness than they would if they were not flattened. In the plot, each curve spans 360° of rotation for a different orientation of the pole. Dotted lines show the intermediate light curves at 15° steps in $\phi$. The inset figures show each shape's appearance on the sky at minimum brightness ($\theta = 90°$), for the specified value of $\phi$ (rotated toward the right).

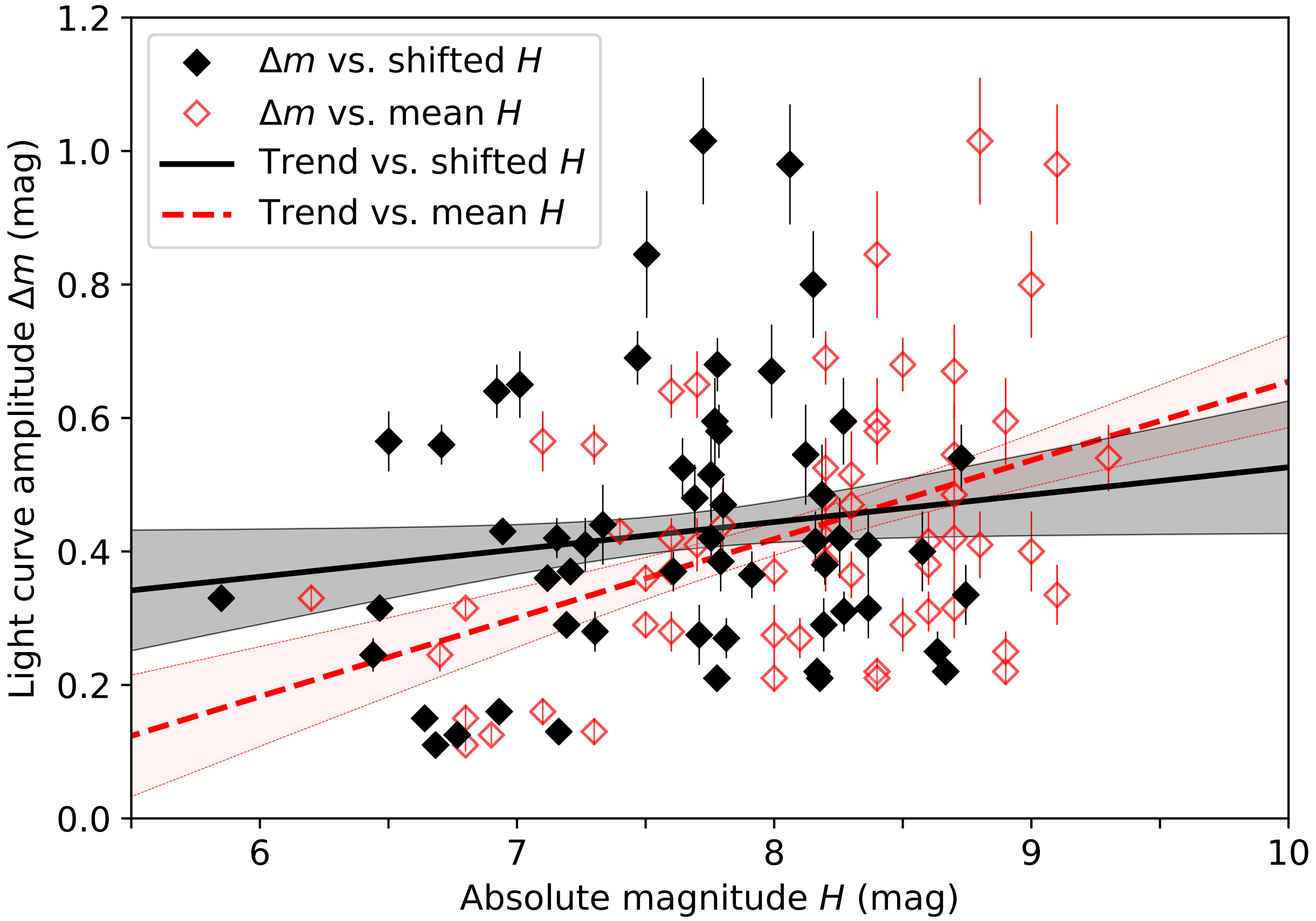

**Figure 14.** Implications of flattening on the observed trend of light curve amplitudes $\Delta m$ versus absolute magnitude $H$. Open symbols show $\Delta m$ versus mean $H$, as they do in Fig. 12. Solid symbols show the same measurements, but here, each abscissa $H$ has been shifted by -1.06 $\Delta m$, an amount appropriate to compensate for bodies flattened by a factor $c/b$ = 0.6. The measurements are from the OSSOS data set (Alexandersen et al., 2019), matching those shown in Fig. 12c. As in Fig. 12, each set of points is fitted to a linear trend, using a dashed line for the fit to open-symbol measurements and a solid line for the fit to solid-symbol measurements. The shaded regions around each line illustrate the ±1σ uncertainty in each fit. Although the dashed line shows a statistically significant trend in $\Delta m$ with mean $H$ (3.5σ significance; *p*-value = $2.4 \times 10^{-4}$), that trend is much less significant when fitted to the shifted measurements (1.0σ; *p*-value = 0.15). This figure illustrates how a population of flattened bodies, all having identical proportions, could give the false impression that smaller bodies are more irregular than large ones.

This makes it worth exploring the possible role of flattening in any observed dependence of $\Delta m$ on size. In this case, $H_{min}$ is no longer a good proxy for size, although it is still superior to $H_0$.

The ideal proxy can be written $H' = H_0 - C\,\Delta m$, but the proportionality constant $C$ depends on the (unknown) shapes of the bodies. For ellipsoids, the general expression is $C \approx 0.5 - \log(c/b) / \log(a/b)$. Note that $C$ can be quite large if $a \approx b$, because $\log(a/b)$ in the denominator is small. However, such bodies also have small $\Delta m$, so they make minimal contributions to any observed trend in $\Delta m$ vs. size.

As a simple test of how flattening might play a role in observed trends, let us suppose that all the bodies in the OSSOS data set (Fig. 12c) are flattened by a factor $c/b = 0.6$, similar to the two lobes of Arrokoth (Spencer et al., 2020). Although we cannot know $a/b$, a conservative choice would be a relatively large ratio, because this will reduce $C$ and minimize the distances separating $H_0$ and $H'$ on the plot. Because the largest amplitudes in the OSSOS photometry are ~ 1 magnitude, we adopt $a/b = 2.5$, yielding $C = 1.06$. Figure 14 shows the result of this test. The slope of the trend line—the extent to which $\Delta m$ really depends on the sizes of KBOs—is now 0.041 ± 0.040 ($p$-value = 0.15), providing a much lower level of confidence than before.

The purpose of this exercise has not been to prove or disprove any particular hypothesis about any one data set, or about the flattening of KBOs. Instead, we have merely highlighted some of the subtle assumptions that go into inferences about the relationship between “size” and light curve amplitude. In particular, we have shown that an ensemble of flattened bodies with fixed proportions, independent of size, can produce a strong, positive correlation between $H$ and $\Delta m$. This may seem counterintuitive, but it occurs because bodies with higher $\Delta m$ also have values of $H$ that have been systematically shifted rightward in the plots. In the end, it may still be true that smaller KBOs are more irregular than large ones—such a finding would not be all that surprising—but we will need more data and further analysis to determine if this is the case.

## 5. Discussion

### *5.1 OSSOS and non-OSSOS Photometry*

As we have discussed, it is essentially impossible to understand how the OSSOS and non-OSSOS photometry could be representative of the same KBO population. We have shown that, in several analyses, these two data sets lead to fundamentally incompatible conclusions. Because the OSSOS data set has well-understood selection biases (Bannister et al., 2016), and was processed using a single calibration pipeline (Alexandersen et al., 2019), we would have expected it to be ideal for our purposes. However, the statistical properties of the data remain especially challenging to understand. Most notable is the dearth of low-amplitude light curves, which would imply that the obliquities of KBOs are preferentially clustered around 0° and/or 180° (Fig. 11). If true, this result would serve as an important constraint on KBO formation and evolution scenarios (e.g., Nesvorný et al., 2019).

However, a straightforward inspection of the individual photometric measurements (Figs. 2–6 of Alexandersen et al., 2019) reveals that adjacent measurements, separated by only ~ 36 minutes, often differ by considerably more than their reported uncertainties. This raises a concern that, at minimum, the error bars might have been underestimated. We can quantify this general impression as follows. A sequence of five consecutive measurements in this data set spans roughly 2.4 hours. Within that interval, a body with a period ≳ 5 hours, having a double-peaked light curve, would show no more than two extrema. Slower rotators would generally exhibit slower and more uniform variations. For comparison, reported rotation periods for KBOs are generally longer than 6 hours, and often several times longer (Alexandersen et al., 2019; Thirouin and Sheppard, 2018, 2019). Thus, to within the photometric precision of the Alexandersen et al. data set, we would expect almost any 2.4-hour segment of a light curve to show smooth variations that are reasonably well described by a cubic equation (which can have up to two extrema). We therefore define a metric $Q$ as the RMS deviation of each measurement from the cubic curve fitted to its four nearest neighbors. Numerical experiments show that, when all five measurements are timed uniformly and have similar uncertainties $\sigma$, the expected value of $Q$ is $1.4\sigma$. For comparison, Alexandersen et al.'s photometry contains 107 independent, 5-sample subsets that each span ≤ 2.5 hours. Among these sequences, we obtain $Q = 4.0\sigma$, suggesting that OSSOS uncertainties are systematically small by a factor of ~ 3. When we use the K-S statistic to compare the 107 measurements described above to simulated data, the likelihood that they represent the same distribution is $\sim 2 \times 10^{-10}$.

Overlooking the issue described above, one might try to reconcile the two data sets by noting that the OSSOS data set describes generally smaller bodies (cf. Figs. 12b,c). However, if we limit our selections to KBOs with $H \leq 8$, where the two data sets largely overlap, the distinctions persist. The K-S test tells us that we can still reject the hypothesis that the data sets sample the same $\Delta m$ distribution with 97.3% confidence ($p$-value = 0.027).

Because of these test results, combined with the fundamental incompatibility of the OSSOS measurements with any of the models shown in Fig. 11, our preferred interpretation is that the OSSOS calibration pipeline has errors of unknown origin. Clearly, this issue warrants further study.

### *5.2 Pole Distributions*

Our analysis highlights the role of the pole distribution in understanding KBO light curves. While knowing the mutual orbital inclinations of KBO binaries (cf. Grundy et al., 2019) is extremely useful, it is unclear that the angular momentum vectors of binaries and singles should have similar directional distributions.

Direct constraints on the pole directions can be derived from changes in a light curve as a KBO orbits the Sun, but these orbits of course take centuries. Lacerda (2011) and Fernandez-Valenzuela et al. (2019) have demonstrated, however, that time scales of ten years or less may be sufficient to obtain useful pole constraints for some KBOs. However, here again, we need to be aware of a strong selection bias. Over time scales that are very short compared to the orbital period, changes will only be detectable for KBOs with the largest time derivatives, $d(\Delta m)/dt$. By the chain rule, this equals $d(\Delta m)/d\phi \cdot d\phi/dt$. The second term is largest for bodies with obliquities near 90° (similar to Uranus), and/or with $\phi$ near 90° (i.e., currently near their equinox). The first term, d(Δ$m$)/dϕ, is largest for the most elongated bodies and, as Fig. 2a demonstrates, can be especially steep for the “V”-shaped light curves of contact binaries.

Thus, those KBOs whose poles can be constrained with a few decades of Earth-based data will never be representative of the full population, either in terms of shape or orientation. To remove this source of bias in our data, we would require measurements that span one or more orbits of each KBO—in other words, centuries. In the nearer term, we will need to find ways to de-bias any observational constraints on the distribution of KBO rotation poles.

### *5.3 Summary*

We have discussed how the statistical properties of light curves relate to the shapes and orientations of KBO populations. Our analysis is similar to that of Masiero et al. (2009), who applied a similar statistical analysis to a database of 828 main belt asteroids. Their work revealed the general distribution of asteroid shapes when modeled as triaxial ellipsoids. Although no comparable data set exists for KBOs today, the upcoming Vera C. Rubin Observatory, which will be conducting the Legacy Survey of Space and Time (LSST) is expected to increase the number of known KBOs to ~ 30,000 and will observe each one ~ 800 times over a period of ten years (Ivezić et al., 2019). The timing of these measurements will not necessarily support the determinations of KBO rotation periods, but the very large numbers of measurements will allow us to sample the amplitude distribution of many KBOs with very fine precision. Any analysis of the LSST data set will be necessarily statistical, and the framework we have presented here will help to make that possible.

As we have shown (Fig. 11), the cumulative distribution of light curve amplitudes depends on two unknowns, shape and pole orientation. This makes it difficult to draw firm conclusions about one without making assumptions about the other. However, additional information will eventually be at our disposal to break these degeneracies. For example, if we were to find that essentially all high-amplitude light curves of small KBOs show the characteristic broad brightness peaks and narrow minima suggestive of contact binaries, then we could rule out ellipsoid models entirely. This would remove one major source of uncertainty in this analysis; in effect, we could eliminate four of the eight sets of models shown in Fig. 11.

In this context, it is worth noting that previous estimates of the “contact binary fraction” in the Kuiper Belt are 10%–25% for cold classicals and up to 50% for Plutinos (Sheppard and Jewitt 2004; Thirouin and Sheppard 2018, 2019). However, these numbers do not always fully account for dependencies on the pole distribution. For comparison, our statistical analysis of the available photometry, limited to single KBOs with $H \geq 5.5$, identifies several models for shape and orientation in which the contact binary fraction could still be 100%. We find it interesting that the simplest possible set of assumptions, involving equal-sized, un-flattened, contact binaries with an isotropic distribution of poles (blue curves in Fig. 11b) provides an extremely good match to the data. We do not advocate this point of view, however, because many other interpretations of the data are possible and, of course, Arrokoth does not fit this description. Nevertheless, our analysis suggests that once one fully accounts for the directional distribution of KBO rotation poles, the fraction of contact binaries in the Kuiper Belt is likely to be higher than in previous estimates.

Although it is perfectly natural for astronomers to focus on those high-amplitude light curves that can be modeled individually, those with low amplitude should also be pursued observationally and should be consistently reported in the literature. Although it may seem counterintuitive, these low-amplitude light curves also contain fundamental information about the properties of the Kuiper Belt.

## Data Availability

The software library used to generate the 3-D shapes and light curve models appearing in this article was written by lead author Showalter and is permanently archived at https://dmp.seti.org/mshowalter/lightcurves/. Additional documentation, sample programs and data files are also provided.

## Acknowledgments

This work was funded by NASA's New Horizons project. Coauthor Benecchi also acknowledges support for this work under NASA Grant/Contract/Agreement No. NNX15AE04G issued through the SSO Planetary Astronomy Program.

### *Author contributions*

This work emerged from the approach-phase observations for the New Horizons flyby of (486958) 2014 $MU_{69}$ Arrokoth; all the named coauthors participated in the planning, analysis and scientific discussions surrounding this activity.

### *Competing interests*

The authors have no competing interests to declare.

## Appendix A. Analytic Models for Light Curves

As discussed above (Section 2.1), the surface reflectivity $R$ of a body is constant for a Lommel-Seeliger law (Lumme and Bowell, 1981) in the limit of small phase angle. Here it becomes possible to express light curve models for ellipsoids and touching spheres analytically.

The projected area $A$ of an ellipsoid is:

$$A = \pi \left[(a\, c \cos\theta \sin\phi)^2 + (b\, c \sin\theta \sin\phi)^2 + (a\, b \cos\phi)^2\right]^{1/2} . \qquad \text{(A1)}$$

where $a \geq b \geq c$ are the radii along the three principal axes, $\theta$ is the rotation angle about the axis (starting from the long axis), and $\phi$ is the angle between the axis and the line of sight.

For a pair of touching spheres of radius $r_1$ and $r_2$, where $r_1 \geq r_2$, the projected area $A$ can be calculated in a few steps. The projected separation distance between the centers of the two spheres, $d$, can be determined from:

$$d = (r_1 + r_2)\,(\cos^2\theta + \sin^2\theta \cos^2\phi)^{1/2}. \qquad \text{(A2)}$$

We can then determine the half-angles $\xi_1$ and $\xi_2$ for the sector inside each circle within which it overlaps the other circle:

$$\cos \xi_1 = (d^2 + r_1^2 - r_2^2) / (2\, d\, r_1) \; ; \qquad \text{(A3)}$$

$$\cos \xi 2 = (d^2 + r_2^2 - r_1^2) / (2\, d\, r_2) \; . \qquad \text{(A4)}$$

If $\cos \xi_2 < -1$, then the smaller of the two spheres falls entirely in front of or behind the larger one, so $A = \pi\, r_1^2$. Otherwise,

$$A = r_1^2 (\pi - \xi_1 + \sin \xi_1 \cos \xi 1) + r_2^2 (\pi - \xi_2 + \sin \xi_2 \cos \xi_2) \; . \qquad \text{(A5)}$$

# Table 1: KBO Photometry

| Object | Number | Name | Class | Δ*m* (mag) | uncertainty (mag) | min (mag) | max (mag) | *H* (mag) | Ref | Weight | Note |
|---|---|---|---|---|---|---|---|---|---|---|---|
| | 134340 | Pluto | res | | | | | -0.7 | | 0 | H < 5.5, multiple bodies |
| 1977 UB | 2060 | Chiron | cen | = 0.09 | ± 0.01 | 0.08 | 0.10 | 5.8 | TB06 | 1 | |
| 1992 AD | 5145 | Pholus | cen | = 0.15 | | 0.15 | 0.15 | 7.1 | RT99 | 1 | |
| 1993 HA$_2$ | 7066 | Nessus | cen | < 0.20 | | 0.00 | 0.20 | 9.6 | TB06 | 1 | |
| 1993 SC | 15789 | | res | | | | | 7.0 | TS18 | 0 | inconsistent photometry |
| 1994 TB | 15820 | | res | | | | | 7.3 | TS18 | 0 | inconsistent photometry |
| 1994 VK$_8$ | 19255 | | cla | = 0.42 | | 0.42 | 0.42 | 7.0 | RT99 | 1 | |
| 1995 DW$_2$ | 10370 | Hylonome | cen | < 0.04 | | 0.00 | 0.04 | 8.6 | RT99 | 1 | |
| 1995 GO | 8405 | Asbolus | cen | = 0.55 | | 0.55 | 0.55 | 9.1 | S08 | 1 | |
| 1995 HM$_5$ | | | res | > 0.10 | | | | 7.7 | TS18 | 0 | one night |
| 1995 QY$_9$ | 32929 | | res | = 0.60 | ± 0.04 | 0.56 | 0.64 | 8.4 | TS18 | 1 | |
| 1995 SM$_{55}$ | 24835 | | sca | = 0.19 | ± 0.05 | | | 4.6 | T16 | 0 | H < 5.5 |
| 1996 GQ$_{21}$ | 26181 | | sca | < 0.10 | | | | 4.9 | SJ02 | 0 | H < 5.5 |
| 1996 TL$_{66}$ | 15874 | | sca | = 0.07 | ± 0.02 | | | 5.3 | T10 | 0 | H < 5.5 |
| 1996 TO$_{66}$ | 19308 | | sca | = 0.26 | ± 0.03 | | | 4.8 | T16 | 0 | H < 5.5 |
| 1996 TP$_{66}$ | 15875 | | res | < 0.12 | | 0.00 | 0.12 | 7.0 | TS18 | 1 | |
| 1996 TQ$_{66}$ | 118228 | | res | < 0.22 | | 0.00 | 0.22 | 6.9 | TS18 | 1 | |
| 1996 TS$_{66}$ | | | cla | < 0.16 | | 0.00 | 0.16 | 6.1 | TB06 | 1 | |
| 1997 CQ$_{29}$ | 58534 | Logos-Zoe | cla | > 0.50 | | | | 6.6 | TS19 | 0 | known binary |
| 1997 CS$_{29}$ | 79360 | Sila-Nunam | cla | = 0.12 | ± 0.01 | | | 5.3 | R14 | 0 | H < 5.5, known binary |
| 1997 CU$_{26}$ | 10199 | Chariklo | cen | = 0.11 | | 0.11 | 0.11 | 6.7 | F14 | 1 | |
| 1997 CV$_{29}$ | 523899 | | cla | = 0.40 | ± 0.10 | 0.30 | 0.50 | 7.3 | CK04 | 1 | |
| 1998 BU$_{48}$ | 33128 | | cen | = 0.68 | ± 0.04 | 0.64 | 0.72 | 6.7 | SJ02 | 1 | |
| 1998 HK$_{151}$ | 91133 | | res | < 0.15 | | 0.00 | 0.15 | 7.6 | TS18 | 1 | |
| 1998 SG$_{35}$ | 52872 | Okyrhoe | cen | = 0.07 | ± 0.01 | 0.06 | 0.08 | 10.8 | T10 | 1 | |
| 1998 SM$_{165}$ | 26308 | | res | = 0.56 | ± 0.03 | | | 5.7 | TS18 | 0 | known binary |
| 1998 SN$_{165}$ | 35671 | | cla | = 0.15 | ± 0.01 | 0.14 | 0.16 | 5.5 | TB06 | 1 | |
| 1998 VG$_{44}$ | 33340 | | res | < 0.10 | | 0.00 | 0.10 | 6.5 | TS18 | 1 | |
| 1998 WH$_{24}$ | 19521 | Chaos | cla | < 0.10 | | | | 4.8 | TB06 | 0 | H < 5.5 |
| 1998 XY$_{95}$ | 523965 | | sca | < 0.10 | | 0.00 | 0.10 | 6.4 | TB06 | 1 | |
| 1999 CD$_{158}$ | 469306 | | sca | = 0.49 | ± 0.03 | | | 5.0 | T16 | 0 | H < 5.5 |
| 1999 DE$_9$ | 26375 | | res | < 0.10 | | | | 4.8 | S08 | 0 | H < 5.5 |
| 1999 DF$_9$ | 79983 | | cla | = 0.40 | ± 0.02 | 0.38 | 0.42 | 5.8 | S08 | 1 | |
| 1999 KR$_{16}$ | 40314 | | sca | = 0.12 | ± 0.06 | 0.06 | 0.18 | 5.5 | T16 | 1 | |
| 1999 OX$_3$ | 44594 | | cen | = 0.11 | ± 0.02 | 0.09 | 0.13 | 7.0 | T12 | 1 | |
| 1999 OY$_3$ | 86047 | | sca | = 0.08 | ± 0.02 | 0.06 | 0.10 | 6.8 | T16 | 1 | |
| 1999 RZ$_{253}$ | 66652 | Borasisi-Pabua | cla | = 0.08 | ± 0.02 | | | 5.9 | TS19 | 0 | known binary |
| 1999 TC$_{36}$ | 47171 | Lempo | res | < 0.10 | | | | 4.8 | TS18 | 0 | H < 5.5, known triple |
| 1999 TD$_{10}$ | 29981 | | cen | = 0.65 | ± 0.05 | 0.60 | 0.70 | 8.7 | S08 | 1 | |
| 1999 UG$_5$ | 31824 | Elatus | cen | = 0.17 | ± 0.07 | 0.10 | 0.24 | 10.1 | S08 | 1 | |
| 2000 CG$_{105}$ | | | sca | = 0.45 | | 0.45 | 0.45 | 6.6 | T16 | 1 | |
| 2000 CL$_{104}$ | | | cla | > 0.20 | | 0.20 | 0.20 | 6.2 | TS19 | 1 | 2 nights |
| 2000 CM$_{105}$ | 80806 | | cla | < 0.14 | | | | 6.6 | TS19 | 0 | known binary |
| 2000 EB$_{173}$ | 38628 | Huya | res | = 0.02 | ± 0.01 | | | 4.8 | TS18 | 0 | H < 5.5, known binary |
| 2000 EC$_{98}$ | 60558 | Echeclus | cen | = 0.24 | ± 0.06 | | | 9.5 | S08, D14 | 0 | observed activity |
| 2000 FV$_{53}$ | | | res | = 0.07 | ± 0.02 | 0.05 | 0.09 | 8.3 | TS18 | 1 | |
| 2000 GN$_{171}$ | 47932 | | res | = 0.53 | | 0.53 | 0.53 | 6.2 | TS18 | 1 | |
| 2000 OK$_{67}$ | 138537 | | cla | > 0.15 | | 0.15 | 0.15 | 6.2 | TS19 | 1 | 4 nights |
| 2000 OU$_{69}$ | | | cla | > 0.15 | | | | 6.9 | TS19 | 0 | one night |
| 2000 QA$_{243}$ | | | res | = 0.16 | + 0.02 - 0.02 | 0.14 | 0.18 | 7.1 | A19 | 1 | |
| 2000 QC$_{243}$ | 54598 | Bienor | cen | = 0.75 | ± 0.09 | 0.66 | 0.84 | 7.5 | S08 | 1 | |
| 2000 QH$_{226}$ | | | res | = 0.48 | + 0.06 - 0.06 | | | 9.0 | A19 | 0 | one night |
| 2000 WR$_{106}$ | 20000 | Varuna | sca | = 0.43 | ± 0.01 | | | 3.6 | T10 | 0 | H < 5.5 |
| 2000 YW$_{134}$ | 82075 | | res | < 0.10 | | | | 4.5 | S08 | 0 | H < 5.5, known binary |
| 2001 CZ$_{31}$ | 150642 | | cla | = 0.21 | ± 0.02 | 0.19 | 0.23 | 5.9 | S08 | 1 | |
| 2001 FP$_{185}$ | 82158 | | sca | < 0.06 | | 0.00 | 0.06 | 6.2 | S08 | 1 | |
| 2001 FZ$_{173}$ | 82155 | | cen | < 0.06 | | 0.00 | 0.06 | 6.1 | S08 | 1 | |
| 2001 KB$_{77}$ | 469362 | | res | > 0.15 | | | | 7.4 | TS18 | 0 | one night |
| 2001 KD$_{77}$ | | | res | < 0.07 | | 0.00 | 0.07 | 6.0 | TS18 | 1 | |
| 2001 KX$_{76}$ | 28978 | Ixion | res | = 0.06 | ± 0.03 | | | 3.6 | TS18 | 0 | H < 5.5 |
| 2001 PT$_{13}$ | 32532 | Thereus | cen | = 0.16 | ± 0.02 | 0.14 | 0.18 | 9.1 | S08 | 1 | |
| 2001 QC$_{298}$ | | | sca | = 0.40 | | | | 6.8 | T16 | 0 | known binary |
| 2001 QF$_{298}$ | 469372 | | res | ≈ 0.11 | | | | 5.2 | TS18 | 0 | H < 5.5 |
| 2001 QG$_{298}$ | 139775 | | res | = 1.14 | ± 0.04 | | | 6.8 | TS18 | 0 | known binary |
| 2001 QS$_{322}$ | | | cla | > 0.30 | | | | 7.2 | TS19 | 0 | one night |
| 2001 QT$_{297}$ | 88611 | Teharonhiawako | cla | < 0.15 | | | | 5.8 | TS19 | 0 | known binary |
| 2001 QY$_{297}$ | 275809 | | cla | = 0.49 | ± 0.03 | | | 5.4 | T12 | 0 | H < 5.5, known binary |
| 2001 UQ$_{18}$ | 148780 | Altjira | cla | < 0.30 | | | | 5.7 | T14 | 0 | known binary |
| 2001 UR$_{163}$ | 42301 | | res | < 0.08 | | | | 4.1 | TB06 | 0 | H < 5.5 |
| 2001 XA$_{255}$ | 148975 | | cen | = 0.13 | | 0.13 | 0.13 | 11.1 | T13 | 1 | |

| Object | Number | Name | Class | Δ*m* (mag) | uncertainty (mag) | min (mag) | max (mag) | *H* (mag) | Ref | Weight | Note |
|---|---|---|---|---|---|---|---|---|---|---|---|
| 2001 YH$_{140}$ | 126154 | | res | = 0.13 | ± 0.05 | 0.08 | 0.18 | 5.5 | T10 | 1 | |
| 2002 AW$_{197}$ | 55565 | | sca | = 0.02 | ± 0.02 | | | 3.3 | T16 | 0 | H < 5.5 |
| 2002 CC$_{249}$ | 126719 | | sca | = 0.79 | ± 0.04 | 0.75 | 0.83 | 6.6 | TS17 | 1 | |
| 2002 CR$_{46}$ | 42355 | Typhon | cen | = 0.07 | ± 0.01 | | | 7.6 | T10 | 0 | known binary |
| 2002 GB$_{10}$ | 55576 | Amycus | cen | = 0.16 | ± 0.01 | 0.15 | 0.17 | 7.8 | T10 | 1 | |
| 2002 GH$_{32}$ | | | sca | = 0.36 | ± 0.02 | 0.34 | 0.38 | 6.7 | T16 | 1 | |
| 2002 GO$_{9}$ | 83982 | Crantor | cen | = 0.14 | ± 0.04 | 0.10 | 0.18 | 8.8 | S08 | 1 | |
| 2002 GV$_{31}$ | 469438 | | cla | = 0.35 | ± 0.06 | 0.29 | 0.41 | 6.4 | P15 | 1 | |
| 2002 GZ$_{32}$ | 95626 | | cen | = 0.15 | ± 0.03 | 0.12 | 0.18 | 6.9 | D08 | 1 | |
| 2002 KW$_{14}$ | 307251 | | sca | = 0.25 | ± 0.03 | 0.22 | 0.28 | 5.6 | BS13 | 1 | |
| 2002 KX$_{14}$ | 119951 | | cla | < 0.05 | | | | 4.7 | BS13 | 0 | H < 5.5 |
| 2002 KY$_{14}$ | 250112 | (2007 UL126) | cen | = 0.090 | ± 0.006 | 0.084 | 0.096 | 9.7 | M20 | 1 | |
| 2002 LM$_{60}$ | 50000 | Quaoar | cla | | | | | 2.4 | | 0 | H < 5.5, known binary |
| 2002 PN$_{34}$ | 73480 | | cen | = 0.18 | ± 0.04 | 0.14 | 0.22 | 8.5 | S08 | 1 | |
| 2002 PQ$_{145}$ | 363330 | | cla | ≈ 0.10 | | 0.10 | 0.10 | 5.5 | TS19 | 1 | |
| 2002 TC$_{302}$ | 84522 | | res | = 0.04 | ± 0.01 | | | 3.9 | T12 | 0 | H < 5.5 |
| 2002 TX$_{300}$ | 55636 | | sca | = 0.05 | ± 0.01 | | | 3.4 | T12 | 0 | H < 5.5 |
| 2002 UX$_{25}$ | 55637 | | sca | | | | | 3.7 | T14 | 0 | H < 5.5, known binary, inconsistent photometry |
| 2002 VE$_{95}$ | 55638 | | res | = 0.05 | ± 0.01 | | | 5.3 | TS18 | 0 | H < 5.5 |
| 2002 VS$_{130}$ | 149348 | | cla | ≈ 0.10 | | 0.10 | 0.10 | 6.3 | TS19 | 1 | |
| 2002 VT$_{130}$ | 508869 | | cla | = 0.21 | | | | 5.7 | T14 | 0 | known binary |
| 2002 WC$_{19}$ | 119979 | | res | < 0.03 | | | | 4.7 | BS13 | 0 | H < 5.5, known binary |
| 2003 AZ$_{84}$ | 208996 | | res | = 0.07 | ± 0.01 | | | 3.6 | TS18 | 0 | H < 5.5, known binary |
| 2003 BF$_{91}$ | | | unk | = 1.09 | ± 0.25 | 0.84 | 1.34 | 11.7 | TB06 | 1 | |
| 2003 BG$_{91}$ | | | unk | = 0.18 | ± 0.08 | 0.11 | 0.26 | 10.7 | TB06 | 1 | |
| 2003 BH$_{91}$ | | | unk | < 0.15 | | 0.00 | 0.15 | 11.9 | TB06 | 1 | |
| 2003 CO$_{1}$ | 120061 | | cen | = 0.085 | ± 0.015 | 0.07 | 0.10 | 8.9 | D14 | 1 | |
| 2003 EL$_{61}$ | 136108 | Haumea | sca | | | | | 0.2 | | 0 | H < 5.5, known triple |
| 2003 FE$_{128}$ | 469505 | | res | = 0.50 | ± 0.14 | | | 6.3 | T14 | 0 | known binary |
| 2003 FM$_{127}$ | | | cla | = 0.46 | ± 0.04 | 0.42 | 0.50 | 7.1 | TS19 | 1 | |
| 2003 FX$_{128}$ | 65489 | Ceto | cen | = 0.13 | ± 0.02 | | | 6.4 | D08 | 0 | known binary |
| 2003 FY$_{128}$ | 120132 | | sca | = 0.15 | ± 0.01 | | | 4.6 | T10 | 0 | H < 5.5 |
| 2003 HA$_{57}$ | | | res | = 0.31 | ± 0.03 | 0.28 | 0.34 | 8.1 | TS18 | 1 | |
| 2003 HX$_{56}$ | | | sca | ≈ 0.40 | | 0.40 | 0.40 | 7.1 | T16 | 1 | |
| 2003 MW$_{12}$ | 174567 | Varda | sca | < 0.04 | | | | 3.4 | BS13 | 0 | H < 5.5, known binary |
| 2003 OP$_{32}$ | 120178 | | sca | = 0.18 | ± 0.01 | | | 4.0 | T16 | 0 | H < 5.5 |
| 2003 QE$_{112}$ | | | cla | ≈ 0.10 | | 0.10 | 0.10 | 6.6 | TS19 | 1 | |
| 2003 QJ$_{91}$ | | | cla | > 0.20 | | | | 6.7 | TS19 | 0 | one night |
| 2003 QY$_{111}$ | | | cla | > 0.20 | | | | 6.9 | TS19 | 0 | amplitude varies |
| 2003 QY$_{90}$ | | | cla | = 0.34 | ± 0.06 | | | 6.4 | T14 | 0 | known binary |
| 2003 SN$_{317}$ | | | cla | ≈ 0.10 | | | | 6.5 | TS19 | 0 | one night |
| 2003 SP$_{317}$ | | | cla | = 0.56 | + 0.05 - 0.04 | 0.52 | 0.61 | 7.1 | A19 | 1 | |
| 2003 SQ$_{317}$ | | | sca | = 0.85 | ± 0.05 | 0.80 | 0.90 | 6.6 | T16 | 1 | |
| 2003 TH$_{58}$ | | | res | ≈ 0.20 | | 0.20 | 0.20 | 7.1 | T16 | 1 | |
| 2003 UB$_{313}$ | 136199 | Eris | sca | | | | | -1.1 | | 0 | H < 5.5, known binary |
| 2003 UZ$_{117}$ | 416400 | | sca | = 0.09 | ± 0.01 | | | 5.1 | T16 | 0 | H < 5.5 |
| 2003 UZ$_{413}$ | 455502 | | res | = 0.13 | ± 0.03 | | | 4.3 | TS18 | 0 | H < 5.5 |
| 2003 VB$_{12}$ | 90377 | Sedna | sca | | | | | 1.3 | | 0 | H < 5.5 |
| 2003 VS$_{2}$ | 84922 | | res | = 0.22 | ± 0.01 | | | 4.2 | TS18 | 0 | H < 5.5 |
| 2003 WL$_{7}$ | 136204 | | cen | = 0.05 | | 0.05 | 0.05 | 8.8 | D14 | 1 | |
| 2003 YU$_{179}$ | | | cla | > 0.20 | | | | 6.5 | TS19 | 0 | one night, known binary |
| 2004 DW | 90482 | Orcus | res | | | | | 2.2 | | 0 | H < 5.5, known binary |
| 2004 EU$_{95}$ | 444018 | | cla | ≈ 0.10 | | 0.10 | 0.10 | 7.0 | TS19 | 1 | |
| 2004 GV$_{9}$ | 90568 | | sca | = 0.16 | ± 0.03 | | | 3.8 | S07, D08 | 0 | H < 5.5 |
| 2004 HD$_{79}$ | | | cla | > 0.15 | | | | 5.7 | TS19 | 0 | known binary |
| 2004 HF$_{79}$ | 469610 | | cla | ≈ 0.15 | | | | 6.3 | TS19 | 0 | known binary |
| 2004 HJ$_{79}$ | 444025 | | cla | > 0.20 | | 0.20 | 0.20 | 6.9 | TS19 | 1 | 3 nights |
| 2004 HP$_{79}$ | | | res | > 0.15 | | | | 6.6 | TS19 | 0 | one night |
| 2004 MT$_{8}$ | | | cla | > 0.20 | | 0.20 | 0.20 | 6.5 | TS19 | 1 | 3 nights |
| 2004 MU$_{8}$ | | | cla | > 0.48 | | | | 6.0 | TS19 | 0 | known binary |
| 2004 NT$_{33}$ | 444030 | | sca | = 0.04 | ± 0.01 | | | 4.7 | T12 | 0 | H < 5.5 |
| 2004 OQ$_{15}$ | | | res | ≈ 0.10 | | | | 6.8 | TS19 | 0 | one night |
| 2004 PT$_{107}$ | 469615 | | sca | = 0.05 | | 0.05 | 0.05 | 5.8 | T16 | 1 | |
| 2004 PV$_{117}$ | | | cla | ≈ 0.10 | | | | 6.5 | TS19 | 0 | known binary |
| 2004 PX$_{107}$ | | | cla | ≈ 0.10 | | | | 7.2 | TS19 | 0 | one night |
| 2004 PY$_{107}$ | | | cla | ≈ 0.10 | | | | 6.5 | TS19 | 0 | one night |
| 2004 SB$_{60}$ | 120347 | Salacia | sca | = 0.06 | ± 0.02 | | | 4.1 | T16 | 0 | H < 5.5, known binary |
| 2004 TT$_{357}$ | | | res | = 0.76 | ± 0.03 | 0.73 | 0.79 | 8.0 | TSN17 | 1 | |
| 2004 TY$_{364}$ | 120348 | | sca | = 0.22 | ± 0.02 | | | 4.3 | S07 | 0 | H < 5.5 |
| 2004 UX$_{10}$ | 144897 | | cla | = 0.08 | ± 0.01 | | | 4.4 | T10 | 0 | H < 5.5 |

| Object | Number | Name | Class | Δ*m* (mag) | uncertainty (mag) | min (mag) | max (mag) | *H* (mag) | Ref | Weight | Note |
|---|---|---|---|---|---|---|---|---|---|---|---|
| 2004 VC$_{131}$ | | | cla | = 0.55 | ± 0.04 | 0.51 | 0.59 | 6.2 | TS19 | 1 | |
| 2004 VU$_{75}$ | | | sca | > 0.42 | | 0.42 | 0.42 | 6.7 | TS19 | 1 | many nights |
| 2004 XA$_{192}$ | 230965 | | sca | = 0.07 | ± 0.02 | | | 4.2 | T12 | 0 | H < 5.5 |
| 2005 CB$_{79}$ | 308193 | | sca | = 0.05 | ± 0.02 | | | 4.6 | T16 | 0 | H < 5.5 |
| 2005 EF$_{298}$ | 469705 | | cla | = 0.31 | ± 0.04 | | | 5.9 | BS13 | 0 | known binary |
| 2005 EX$_{297}$ | 525460 | | cla | ≈ 0.10 | | | | 6.5 | TS19 | 0 | one night |
| 2005 FY$_{9}$ | 136472 | Makemake | sca | | | | | -0.1 | | 0 | H < 5.5, known binary |
| 2005 GE$_{187}$ | 469708 | | res | = 0.29 | ± 0.02 | 0.27 | 0.31 | 7.3 | TS18 | 1 | |
| 2005 JP$_{179}$ | 525595 | | cla | ≈ 0.08 | | | | 6.7 | TS19 | 0 | one night |
| 2005 PL$_{21}$ | | | cla | > 0.15 | | | | 6.6 | TS19 | 0 | one night |
| 2005 PR$_{21}$ | 303712 | | cla | < 0.28 | | | | 6.2 | BS13 | 0 | known binary |
| 2005 QU$_{182}$ | 303775 | | sca | = 0.12 | ± 0.02 | | | 3.6 | BS13 | 0 | H < 5.5 |
| 2005 RM$_{43}$ | 145451 | | sca | = 0.04 | ± 0.01 | | | 4.4 | T10 | 0 | H < 5.5 |
| 2005 RN$_{43}$ | 145452 | | sca | = 0.06 | ± 0.01 | | | 3.7 | BS13 | 0 | H < 5.5 |
| 2005 RR$_{43}$ | 145453 | | sca | < 0.06 | | | | 4.0 | T16 | 0 | H < 5.5 |
| 2005 TB$_{190}$ | 145480 | | sca | = 0.12 | ± 0.01 | | | 4.4 | T12 | 0 | H < 5.5 |
| 2005 UJ$_{438}$ | 145486 | | cen | = 0.13 | | 0.13 | 0.13 | 10.8 | D14 | 1 | |
| 2005 UQ$_{513}$ | 202421 | | sca | = 0.06 | ± 0.02 | | | 3.6 | T16 | 0 | H < 5.5 |
| 2006 HJ$_{123}$ | 469987 | | res | < 0.13 | | 0.00 | 0.13 | 5.9 | BS13 | 1 | |
| 2006 UZ$_{184}$ | | | res | > 0.20 | | 0.20 | 0.20 | 8.1 | TS18 | 1 | 2 nights |
| 2007 JF$_{43}$ | 444745 | | res | = 0.22 | ± 0.02 | | | 5.3 | BS13 | 0 | H < 5.5 |
| 2007 JH$_{43}$ | 470308 | | sca | < 0.08 | | | | 4.5 | BS13 | 0 | H < 5.5 |
| 2007 JJ$_{43}$ | 278361 | | sca | = 0.100 | ± 0.005 | | | 4.5 | P15 | 0 | H < 5.5 |
| 2007 OR$_{10}$ | 225088 | Gongong | res | | | | | 1.6 | | 0 | H < 5.5, known binary |
| 2007 TY$_{430}$ | 341520 | Mors-Somnus | res | = 0.24 | ± 0.05 | | | 6.6 | TS18 | 0 | known binary |
| 2007 UK$_{126}$ | 229762 | G!kúnǁ'hòmdímà | sca | = 0.03 | ± 0.01 | | | 3.3 | T14 | 0 | H < 5.5, known binary |
| 2008 AP$_{129}$ | 315530 | | sca | = 0.12 | ± 0.02 | | | 4.7 | T16 | 0 | H < 5.5 |
| 2008 QD$_{4}$ | 315898 | | cen | ≈ 0.09 | | 0.09 | 0.09 | 11.3 | T13 | 1 | |
| 2008 QY$_{40}$ | 305543 | | sca | < 0.15 | | | | 5.5 | BS13 | 0 | H < 5.5 |
| 2008 YB$_{3}$ | 342842 | | cen | = 0.21 | | 0.21 | 0.21 | 9.3 | PA13 | 1 | |
| 2009 YD$_{7}$ | 353222 | | cen | = 0.21 | ± 0.02 | 0.15 | 0.23 | 9.8 | M20 | 1 | |
| 2009 YE$_{7}$ | 386723 | | sca | ≈ 0.18 | | | | 4.3 | T16 | 0 | H < 5.5 |
| 2010 BK$_{118}$ | | | cen | ≈ 0.15 | | 0.15 | 0.15 | 10.2 | T13 | 1 | |
| 2010 EK$_{139}$ | 471143 | Dziewanna | res | = 0.12 | ± 0.02 | | | 3.9 | BS13 | 0 | H < 5.5 |
| 2010 EL$_{139}$ | | | res | = 0.15 | ± 0.03 | 0.12 | 0.18 | 5.6 | BS13 | 1 | |
| 2010 EP$_{65}$ | 312645 | | res | = 0.17 | ± 0.03 | | | 5.3 | BS13 | 0 | H < 5.5 |
| 2010 ER$_{65}$ | | | sca | < 0.16 | | | | 5.2 | BS13 | 0 | H < 5.5 |
| 2010 ET$_{65}$ | 471137 | | sca | = 0.13 | ± 0.02 | | | 5.1 | BS13 | 0 | H < 5.5 |
| 2010 FX$_{86}$ | | | sca | = 0.26 | ± 0.04 | | | 4.7 | BS13 | 0 | H < 5.5 |
| 2010 GX$_{34}$ | | | cen | < 0.60 | | 0.00 | 0.60 | 8.6 | M20 | 1 | |
| 2010 HE$_{79}$ | 471165 | | sca | = 0.11 | ± 0.02 | | | 5.1 | BS13 | 0 | H < 5.5 |
| 2010 JJ$_{124}$ | | | cen | < 0.50 | | 0.00 | 0.50 | 6.7 | M20 | 1 | |
| 2010 KZ$_{39}$ | | | sca | < 0.17 | | | | 4.0 | BS13 | 0 | H < 5.5 |
| 2010 PL$_{66}$ | 499522 | | cen | < 0.20 | | 0.00 | 0.20 | 7.6 | M20 | 1 | |
| 2010 PU$_{75}$ | | | sca | = 0.27 | ± 0.02 | 0.25 | 0.29 | 6.0 | BS13 | 1 | |
| 2010 RF$_{43}$ | | | sca | < 0.08 | | | | 3.9 | BS13 | 0 | H < 5.5 |
| 2010 RO$_{64}$ | 523640 | | sca | < 0.16 | | | | 5.2 | BS13 | 0 | H < 5.5 |
| 2010 TF$_{192}$ | | | cla | > 0.30 | | | | 6.1 | TS19 | 0 | one night |
| 2010 TL$_{182}$ | | | cla | > 0.25 | | | | 6.5 | TS19 | 0 | one night |
| 2010 TY$_{53}$ | 523643 | | cen | < 0.14 | | 0.00 | 0.14 | 5.7 | BS13 | 1 | |
| 2010 VK$_{201}$ | 523645 | | sca | = 0.30 | ± 0.02 | | | 5.0 | BS13 | 0 | H < 5.5 |
| 2010 VZ$_{98}$ | 445473 | | sca | < 0.18 | | | | 4.8 | BS13 | 0 | H < 5.5 |
| 2011 BV$_{163}$ | 530231 | | cla | > 0.15 | | | | 6.6 | TS19 | 0 | one night |
| 2012 DA$_{99}$ | 531076 | | cla | ≈ 0.10 | | | | 6.8 | TS19 | 0 | one night |
| 2012 DZ$_{98}$ | | | cla | > 0.20 | | | | 6.5 | TS19 | 0 | one night |
| 2012 VU$_{85}$ | 463368 | | cen | = 0.38 | ± 0.05 | 0.33 | 0.43 | 7.3 | M20 | 1 | |
| 2013 AQ$_{183}$ | | | cla | > 0.15 | | 0.15 | 0.15 | 6.8 | TS19 | 1 | 2 nights |
| 2013 EM$_{149}$ | | | cla | ≈ 0.10 | | | | 6.7 | TS19 | 0 | one night |
| 2013 FA$_{28}$ | | | cla | ≈ 0.10 | | 0.10 | 0.10 | 6.2 | TS19 | 1 | |
| 2013 PH$_{44}$ | 471931 | | cen | = 0.15 | ± 0.04 | 0.11 | 0.19 | 9.1 | M20 | 1 | |
| 2013 SM$_{100}$ | | | cla | = 0.68 | + 0.04 - 0.04 | 0.64 | 0.72 | 8.5 | A19 | 1 | |
| 2013 ST$_{102}$ | | | cla | = 0.53 | + 0.04 - 0.05 | 0.48 | 0.57 | 8.2 | A19 | 1 | |
| 2013 UC$_{18}$ | | | cla | = 0.47 | + 0.04 - 0.04 | 0.43 | 0.51 | 8.3 | A19 | 1 | |
| 2013 UK$_{17}$ | | | res | = 0.15 | + 0.02 - 0.02 | 0.13 | 0.17 | 6.8 | A19 | 1 | |
| 2013 UL$_{15}$ | | | cla | = 0.363 | + 0.012 - 0.011 | | | 6.6 | A19 | 0 | known binary |
| 2013 UM$_{15}$ | | | res | = 0.127 | + 0.015 - 0.014 | 0.11 | 0.14 | 6.9 | A19 | 1 | |
| 2013 UN$_{15}$ | | | cla | = 0.56 | + 0.03 - 0.03 | 0.53 | 0.59 | 7.3 | A19 | 1 | |
| 2013 UP$_{15}$ | | | cla | = 0.29 | + 0.02 - 0.02 | 0.27 | 0.31 | 7.5 | A19 | 1 | |
| 2013 UR$_{22}$ | | | cla | = 0.44 | + 0.06 - 0.06 | 0.38 | 0.50 | 7.8 | A19 | 1 | |
| 2013 UT$_{15}$ | | | sca | = 0.33 | + 0.02 - 0.02 | 0.31 | 0.35 | 6.2 | A19 | 1 | |
| 2013 UV$_{17}$ | | | res | = 0.69 | + 0.04 - 0.04 | 0.65 | 0.73 | 8.2 | A19 | 1 | |

| Object | Number | Name | Class | Δ*m* (mag) | uncertainty (mag) | min (mag) | max (mag) | *H* (mag) | Ref | Weight | Note |
|---|---|---|---|---|---|---|---|---|---|---|---|
| 2013 UW$_{16}$ | | | cla | = 0.13 | + 0.02 - 0.02 | 0.11 | 0.15 | 7.3 | A19 | 1 | |
| 2013 UW$_{17}$ | | | cla | = 0.42 | + 0.03 - 0.03 | 0.39 | 0.45 | 7.6 | A19 | 1 | |
| 2013 UX$_{16}$ | | | res | = 0.27 | + 0.05 - 0.04 | 0.23 | 0.32 | 8.0 | A19 | 1 | |
| 2013 UY$_{16}$ | | | cla | = 0.37 | + 0.02 - 0.02 | 0.35 | 0.39 | 7.6 | A19 | 1 | |
| 2013 UZ$_{16}$ | | | cen | = 0.36 | + 0.03 - 0.03 | 0.33 | 0.39 | 7.8 | A19 | 1 | |
| 2014 GZ$_{53}$ | 533397 | | cla | ≈ 0.10 | | | | 6.3 | TS19 | 0 | one night |
| 2014 JK$_{80}$ | | | res | > 0.17 | | 0.17 | 0.17 | 6.4 | TS18 | 1 | 2 nights |
| 2014 JL$_{80}$ | | | res | = 0.55 | ± 0.03 | 0.52 | 0.58 | 7.4 | TS18 | 1 | |
| 2014 JO$_{80}$ | | | res | = 0.60 | ± 0.05 | 0.55 | 0.65 | 7.8 | TS18 | 1 | |
| 2014 JP$_{80}$ | | | res | > 0.10 | | | | 5.0 | TS18 | 0 | H < 5.5 |
| 2014 JQ$_{80}$ | 533562 | | res | = 0.76 | ± 0.04 | 0.72 | 0.80 | 7.3 | TS18 | 1 | |
| 2014 JT$_{80}$ | | | res | > 0.10 | | | | 7.4 | TS18 | 0 | one night |
| 2014 KC$_{102}$ | | | res | > 0.20 | | 0.20 | 0.20 | 7.1 | TS18 | 1 | 2 nights |
| 2014 KX$_{101}$ | | | res | > 0.20 | | 0.20 | 0.20 | 7.5 | TS18 | 1 | 2 nights |
| 2014 LQ$_{28}$ | | | sca | ≈ 0.08 | | | | 5.8 | TS19 | 0 | known binary |
| 2014 LR$_{28}$ | 523721 | | cla | > 0.25 | | | | 5.3 | TS19 | 0 | H < 5.5 |
| 2014 LS$_{28}$ | 533676 | | cla | = 0.35 | | 0.35 | 0.35 | 6.2 | TS19 | 1 | |
| 2014 OA$_{394}$ | | | cla | > 0.15 | | | | 6.8 | TS19 | 0 | one night |
| 2014 OM$_{394}$ | | | cla | ≈ 0.10 | | | | 6.1 | TS19 | 0 | one night |
| 2014 UA$_{225}$ | | | sca | = 0.11 | + 0.01 - 0.01 | 0.10 | 0.12 | 6.8 | A19 | 1 | |
| 2015 BA$_{519}$ | | | res | ≈ 0.16 | | 0.16 | 0.16 | 7.7 | TS18 | 1 | |
| 2015 FZ$_{117}$ | 472760 | | cen | < 0.20 | | 0.00 | 0.20 | 10.6 | M20 | 1 | |
| 2015 RA$_{280}$ | | | cla | = 0.64 | + 0.04 - 0.04 | 0.60 | 0.68 | 7.6 | A19 | 1 | |
| 2015 RB$_{280}$ | | | cla | = 0.55 | + 0.02 - 0.02 | | | 7.6 | A19 | 0 | known binary |
| 2015 RB$_{281}$ | | | cla | = 0.43 | + 0.02 - 0.02 | 0.41 | 0.45 | 7.4 | A19 | 1 | |
| 2015 RC$_{277}$ | | | sca | = 0.37 | + 0.03 - 0.03 | 0.34 | 0.40 | 8.0 | A19 | 1 | |
| 2015 RC$_{280}$ | | | cla | = 0.40 | + 0.06 - 0.06 | 0.34 | 0.46 | 9.0 | A19 | 1 | |
| 2015 RD$_{277}$ | | | cen | = 0.15 | + 0.03 - 0.02 | 0.13 | 0.18 | 10.6 | A19 | 1 | |
| 2015 RD$_{281}$ | | | sca | = 0.54 | + 0.05 - 0.05 | 0.49 | 0.59 | 9.3 | A19 | 1 | |
| 2015 RE$_{278}$ | | | cla | = 0.59 | + 0.07 - 0.06 | 0.53 | 0.66 | 8.9 | A19 | 1 | |
| 2015 RE$_{280}$ | | | cla | = 0.23 | + 0.05 - 0.04 | | | 7.9 | A19 | 0 | one night |
| 2015 RG$_{279}$ | | | sca | = 0.24 | + 0.03 - 0.02 | 0.22 | 0.27 | 6.7 | A19 | 1 | |
| 2015 RG$_{281}$ | | | sca | = 0.21 | + 0.02 - 0.02 | 0.19 | 0.23 | 8.4 | A19 | 1 | |
| 2015 RH$_{279}$ | | | sca | = 0.49 | + 0.07 - 0.08 | 0.41 | 0.56 | 8.7 | A19 | 1 | |
| 2015 RH$_{280}$ | | | cla | = 0.80 | + 0.08 - 0.08 | 0.72 | 0.88 | 9.0 | A19 | 1 | |
| 2015 RH$_{281}$ | | | cla | = 0.59 | + 0.07 - 0.06 | 0.53 | 0.66 | 8.4 | A19 | 1 | |
| 2015 RJ$_{278}$ | | | res | = 0.41 | + 0.04 - 0.04 | 0.37 | 0.45 | 7.7 | A19 | 1 | |
| 2015 RK$_{281}$ | | | cla | = 0.41 | + 0.05 - 0.04 | 0.37 | 0.46 | 8.6 | A19 | 1 | |
| 2015 RL$_{278}$ | | | sca | = 0.41 | + 0.05 - 0.05 | 0.36 | 0.46 | 8.8 | A19 | 1 | |
| 2015 RN$_{278}$ | | | res | = 0.22 | + 0.02 - 0.02 | 0.20 | 0.24 | 8.4 | A19 | 1 | |
| 2015 RN$_{281}$ | | | sca | = 0.51 | + 0.07 - 0.06 | 0.45 | 0.58 | 8.3 | A19 | 1 | |
| 2015 RO$_{278}$ | | | sca | = 0.31 | + 0.03 - 0.03 | 0.28 | 0.34 | 8.6 | A19 | 1 | |
| 2015 RO$_{280}$ | | | sca | = 0.33 | + 0.05 - 0.04 | 0.29 | 0.38 | 9.1 | A19 | 1 | |
| 2015 RO$_{281}$ | | | sca | = 0.36 | + 0.02 - 0.02 | 0.34 | 0.38 | 7.5 | A19 | 1 | |
| 2015 RP$_{281}$ | | | cla | = 0.65 | + 0.05 - 0.05 | 0.60 | 0.70 | 7.7 | A19 | 1 | |
| 2015 RQ$_{280}$ | | | cla | = 1.01 | + 0.10 - 0.09 | 0.92 | 1.11 | 8.8 | A19 | 1 | |
| 2015 RR$_{278}$ | | | res | = 0.40 | + 0.08 - 0.06 | | | 9.7 | A19 | 0 | one night |
| 2015 RR$_{280}$ | | | sca | = 0.27 | + 0.03 - 0.03 | 0.24 | 0.30 | 8.1 | A19 | 1 | |
| 2015 RR$_{281}$ | | | cla | = 0.54 | + 0.08 - 0.07 | 0.47 | 0.62 | 8.7 | A19 | 1 | |
| 2015 RS$_{281}$ | | | res | = 0.98 | + 0.09 - 0.09 | 0.89 | 1.07 | 9.1 | A19 | 1 | |
| 2015 RT$_{277}$ | | | res | = 0.28 | + 0.05 - 0.03 | 0.25 | 0.33 | 8.5 | A19 | 1 | |
| 2015 RT$_{278}$ | | | sca | = 0.38 | + 0.03 - 0.03 | 0.35 | 0.41 | 8.6 | A19 | 1 | |
| 2015 RT$_{279}$ | | | cla | = 0.48 | + 0.05 - 0.05 | 0.43 | 0.53 | 8.2 | A19 | 1 | |
| 2015 RU$_{277}$ | | | res | = 0.22 | + 0.02 - 0.02 | 0.20 | 0.24 | 8.9 | A19 | 1 | |
| 2015 RU$_{278}$ | | | sca | = 0.314 | + 0.014 - 0.014 | 0.30 | 0.33 | 6.8 | A19 | 1 | |
| 2015 RU$_{279}$ | | | cla | = 0.66 | + 0.08 - 0.06 | 0.60 | 0.74 | 8.7 | A19 | 1 | |
| 2015 RV$_{245}$ | | | cen | = 0.10 | + 0.02 - 0.02 | 0.08 | 0.12 | 10.1 | A19 | 1 | |
| 2015 RV$_{277}$ | | | res | = 0.38 | + 0.05 - 0.04 | 0.34 | 0.43 | 8.2 | A19 | 1 | |
| 2015 RV$_{278}$ | | | sca | = 0.58 | + 0.04 - 0.04 | 0.54 | 0.62 | 8.4 | A19 | 1 | |
| 2015 RW$_{278}$ | | | cla | = 0.31 | + 0.05 - 0.04 | 0.27 | 0.36 | 8.7 | A19 | 1 | |
| 2015 RW$_{279}$ | | | cla | = 0.42 | + 0.04 - 0.04 | 0.38 | 0.46 | 8.2 | A19 | 1 | |
| 2015 RW$_{280}$ | | | sca | = 0.84 | + 0.10 - 0.09 | 0.75 | 0.94 | 8.4 | A19 | 1 | |
| 2015 RX$_{279}$ | | | sca | = 0.41 | + 0.07 - 0.05 | 0.36 | 0.48 | 8.7 | A19 | 1 | |
| 2015 RY$_{278}$ | | | cla | = 0.21 | + 0.02 - 0.02 | 0.19 | 0.23 | 8.0 | A19 | 1 | |
| 2015 RY$_{279}$ | | | sca | = 0.36 | + 0.04 - 0.03 | 0.33 | 0.40 | 8.3 | A19 | 1 | |
| 2015 RZ$_{278}$ | | | sca | = 0.25 | + 0.03 - 0.03 | 0.22 | 0.28 | 8.9 | A19 | 1 | |
| 2015 RZ$_{279}$ | | | cla | = 0.28 | + 0.03 - 0.03 | 0.25 | 0.31 | 7.6 | A19 | 1 | |
| 2016 AE$_{193}$ | 514312 | | cen | = 0.228 | ± 0.014 | 0.214 | 0.242 | 8.1 | M20 | 1 | |
| 2017 CX$_{33}$ | 523798 | | cen | = 0.27 | ± 0.11 | 0.16 | 0.38 | 11.2 | M20 | 1 | |

**Table 1**. Photometric data used in this study. We list every KBO and centaur for which we have found low-phase light curve amplitude values in the literature. Class is one of “cla” for classicals, “res” for resonant bodies, “sca” for scattered disk bodies, “cen” for centaurs, and “unk” for bodies with an unknown classification. References are abbreviated as follows: RT99 = Romanishin and Tegler (1999); SJ02 = Sheppard and Jewitt (2002); CK04 = Chorney and Kavelaars (2004); TB06 = Trilling and Bernstein (2006); S07 = Sheppard (2007); D08 = Dotto et al. (2008); S08 = Sheppard et al. (2008); T10 = Thirouin et al. (2010); T12 = Thirouin et al. (2012); BS13 = Benecchi and Sheppard (2013); PA13 = Pinilla-Alonso et al. (2013); T13 = Thirouin (2013); D14 = Duffard et al. (2014); F14 = Fornasier et al. (2014); R14 = Rabinowitz et al. (2014); T14 = Thirouin et al. (2014); P15 = Pál et al. (2015); T16 = Thirouin et al. (2016); TS17 = Thirouin and Sheppard (2017); TSN17 = Thirouin et al. (2017). TS18 = Thirouin and Sheppard (2018); A19 = Alexandersen et al. (2019); TS19 = Thirouin and Sheppard (2019); M20 = Marton et al. (2020). Values of absolute magnitude are from the Minor Planet Center as of August 5, 2020; see https://minorplanetcenter.net/iau/lists/TNOs.html and https://minorplanetcenter.net/iau/lists/Centaurs.html. We omit details for a few dwarf planets and dwarf planet candidates with $H < 3$; due to their very large sizes, these are unambiguously not relevant to this investigation. Columns “min” and “max” are only filled in for weighted bodies.